\documentclass[onecolumn,bibyear]{aa} 
\usepackage{graphicx}
\usepackage{txfonts}
\usepackage{color}
\usepackage[normalem]{ulem}
\usepackage{tabularx}
\usepackage{diagbox}
\usepackage{amsmath}

\usepackage{natbib}
\bibpunct{(}{)}{;}{a}{}{,} 
\usepackage{comment}
\usepackage{bm}
\usepackage[colorlinks=true, linkcolor = blue, urlcolor = magenta, citecolor = blue]{hyperref}

\begin{document}

   \title{A theoretical view of the T-web statistical description of the cosmic web}

   \author{Emma Ayçoberry
          \inst{1}
          \and
          Alexandre Barthelemy
          \inst{2}
          \and
          Sandrine Codis
          \inst{3,1}
          }

   \institute{Sorbonne Université, CNRS, UMR7095, Institut d’Astrophysique de Paris, 98 bis Boulevard Arago, F-75014, Paris, France \\ \email{emma.aycoberry@iap.fr}
    \and Universitäts-Sternwarte, Fakultät für Physik, Ludwig-Maximilians Universität München, Scheinerstr. 1, 81679 München, Germany
        \and Universit\'e Paris-Saclay, Universit\'e Paris Cit\'e, CEA, CNRS, Astrophysique, Instrumentation et Mod\'elisation Paris-Saclay, 91191 Gif-sur-Yvette, France
             }

   \date{Received 5 October 2023 / Accepted 14 March 2024}

  \abstract 
   {The objective classification of the cosmic web into different environments is an important aspect of large-scale structure studies, as it can be used as a tool to study the formation of structures (halos and galaxies) in mode detail, and it forms a link between their properties and the large-scale environment; these different environments also offer another class of objects whose statistics contain cosmological information.}
   {In this paper, we present an analytical framework to compute the probability of the different environments in the cosmic web based on the so-called T-web formalism, which classifies structures into four different classes (voids, walls, filaments, and knots) based on the eigenvalues of the Hessian of the gravitational potential, often called the tidal tensor.}
   {Our classification method relies on studying whether the eigenvalues of this Hessian matrix are above or below a given threshold and thus requires knowledge of the joint probability distribution of those eigenvalues. We performed a change of variables in terms of rotational invariants, which are polynomials of the field variables and minimally correlated. We studied the distribution of those variables in the linear and quasi-linear regimes with the help of a so-called Gram-Charlier expansion, using tree-order Eulerian perturbation theory to compute the Gram-Charlier coefficients. This expansion then allowed us to predict the probability of the four different environments as a function of the chosen threshold and at a given smoothing scale and redshift for the density field. We checked the validity regime of our predictions by comparing those predictions to measurements made in the N-body Quijote simulations.}
   {Working with fields normalised by their linear variance, we find that scaling the threshold value with the non-linear amplitude of fluctuations allows us to capture almost the entire redshift evolution of the probabilities of the four environments, even if we assume that the density field is Gaussian (corresponding to the linear regime of structure formation). We also show that adding mild non-Gaussian corrections with the help of a Gram-Charlier expansion ---hence introducing corrections that depend on third-order cumulants of the field--- provides even greater accuracy, allowing us to obtain very precise predictions for cosmic web abundances up to scales of as small as $\sim$ 5 Mpc/h and redshifts down to z $\sim $ 0.}
   {}


   \maketitle

\vspace{-0.6cm}
\section{Introduction}
We have long since known that the large-scale distribution of matter throughout the Universe is well-approximated by a filamentary structure dubbed the cosmic web \citep{delapparentetal86}. From a theoretical point of view, the first building blocks for a description of the cosmic web date back to the seminal work of \cite{Zeldovich1970} and many others that followed \citep[e.g.][]{Arnold1982, Klypin1983}. Indeed, the so-called Zeldovich approximation, which describes the first-order ballistic trajectories of particles in a Lagrangian description ---predicts the existence of `pancakes' (\textit{i.e.} sheet-like tenuous walls), filaments, and clusters due to the collapse of anisotropic primordial fluctuations through gravitational instabilities in our expanding Universe. In the 90s, \cite{Bond96} and \cite{Klypin1983} showed that this ensemble of objects forms a connected network called the cosmic web based on the correlations imprinted in the primordial fluctuations. Initial peaks led to the formation of clusters  
at the nodes of the web while initial correlation bridges in between later form filaments that lie within walls, which themselves surround nearly empty void regions. 
Such a web classification can be achieved through the eigenvalues of the linear deformation tensor, which states that if all eigenvalues are negative (or below a threshold that is taken here to be zero) then the region is expanding in 3D, thus describing a void region; if all values are positive then the region is contracting, thus describing a knot, with the other two configurations of eigenvalues leading to either walls or filaments. It is possible to classify the cosmic web by means of its tidal field (Hessian of the gravitational potential). Pioneering work in this direction was presented by \cite{doroskevich1970Ap......6..320D}, and later by \citep{1996MNRAS.281...84V, 2012MNRAS.421..296R, 2018JCAP...03..017D,2016arXiv161103619C, Feldbrugge_2018} 

Another classification method was later introduced by \cite{Hahn_2007a}, \cite{Aragon2007a} and \cite{Forero_Romero_2009} using tidal fields, whereby either the Hessian of the gravitational potential is evaluated theoretically in the linear regime, or the non-linear potential is estimated in dark-matter-only numerical simulations. In subsequent years, several authors developed different classification schemes, to improve the detection of these cosmic web structures in various types of data (continuous fields, simulated datasets, point-like galaxy surveys, etc). For instance, \cite{Aragon2010b} used the SpineWeb topological framework to segment the density field, \cite{Shandarin_2011}, \cite{Falck2012} and \cite{Abel_2012} showed how to classify morphological structures using the Lagrangian phase space sheet to count for shell crossings, and various authors have applied techniques from (continuous and then discrete) Morse theory \citep{Colombi_2000} to identify topological structures in the cosmological density field \citep{2008MNRAS.383.1655S,Aragon2007a,2009MNRAS.393..457S,Sousbie2011}. 
Alternatively, \cite{Hoffman_2012} introduced the V-web classification scheme, this time based on the shear of the velocity field, showing that it is able to better resolve smaller structures, which in turn allows for the study of finer dark-matter halo properties.  
An extension of this idea to Lagrangian settings was later proposed by \cite{Fisher_2016}. We note that the velocity divergence and density fields are closely related (e.g. through the mass-conservation equation in the Vlasov-Poisson system) and their statistics are virtually equivalent in the linear regime of structure formation. We refer readers to \cite{2014ApJ...793...58W} and \cite{2014arXiv1411.4117W} for a thorough investigation of the differences between the dynamical and kinematical classifications.

Beyond the strict motivation of wanting to describe the cosmic web from a mathematical point of view, the classification schemes allow us to explore many different environmental effects on the properties of dark-matter halos \citep{Hahn_2007a,Hahn_2007b,Aragon2007b,Aragon2010a,2012MNRAS.427.3320C,Libeskind2012b, Hellwing_2021} and galaxies within them  \citep{Nuza2014,Metuki2014,2017A&A...597A..86P,2018MNRAS.474..547K,2018MNRAS.481.4753C,2023ApJ...950..114H}. Cosmic web classification can also be used to discriminate between cosmological models, as shown for instance by \cite{2009ApJ...696L..10L} or \cite{Biswas_2010} using voids; \cite{2013MNRAS.435..531C} with counts of cosmic web critical points; \cite{Feldbrugge_2019} with Betti numbers; \cite{2018MNRAS.479..973C} using the connectivity of the filaments; \cite{2022A&A...661A.146B} relying on the power spectrum of the various cosmic web environments; and \cite{Dome_2023} with cosmic web abundances.  
Understanding how this cosmic web evolves with time and scale is therefore paramount for both cosmology and studies of the galaxy formation that takes place within this large-scale environment. 
This evolution has been studied theoretically in some contexts (e.g. the local skeleton; \cite{2009MNRAS.396..635P}) with semi-analytical approaches \citep[e.g.][]{Fard_2019} and in various numerical works. One example of the latter is the recent work by \cite{Cui_2017,Cui_2019}, who used simulations to investigate the abundances of the various environments and their time evolution, relying on the T- and V-web decompositions. However, to the best of our knowledge, no theoretical model in the quasi-linear regime, and based on first principles, has been explicitly derived so far for these cosmic web abundances. This objective is nevertheless within the reach of standard techniques used in large-scale structure theoretical studies.

In the present paper, we therefore propose to derive theoretical predictions for web classifications based on the T-web definition. We first focus on a Gaussian description of the cosmic density field before turning to mild non-Gaussian corrections.

To assess the validity regime of our theoretical formalism and predictions for the one-point statistics of all four cosmic environments, we compare our results with measurements made in a dark-matter-only N-body numerical simulation, the Quijote simulation \citep{Quijote_sims}. The simulated box used by these latter authors has a size of $1$ Gpc/h, with $1024^3$ dark matter particles. We are using their high-resolution fiducial cosmology snapshots at five different redshifts: $z=0, 0.5, 1, 2,$ and $3$. Their fiducial cosmology is \{$\Omega_m,~\Omega_b,~h,~n_s, \sigma_8$\} = \{$0.3175,~0.049,~0.6711,~0.9624,~0.834$\}. Finally, in order to maximise the statistical relevance of our analysis, we typically average our measurements over 11 realisations of the Quijote suite and use those realisations to estimate error bars.

The outline of the paper is as follows. We first describe the T-web and V-web classifications in section~\ref{tweb}. In section~\ref{gaussian}, we then present the theoretical formalism and the predictions obtained for the abundance of the different environments (as a function of threshold, redshift, and smoothing scale) in the linear regime of structure formation, that is assuming a Gaussian matter density field. Section~\ref{GC} introduces a Gram-Charlier expansion to include non-Gaussian (i.e. non-linear) corrections to the joint probability distribution function (joint-PDF) of the elements of the Hessian of the gravitational potential, and we finally discuss our results and draw conclusions in section~\ref{conclusion}. In the remainder of this paper, we will use 'abundance' or 'probability' equivalently.

\section{T-web classification of the cosmic web} \label{tweb}

Among many possibilities \citep{Libeskind_2017, Leclercq_2016}, one commonly used mathematical way to classify the different cosmological environments is based on the number of eigenvalues of a deformation tensor above a threshold. Several definitions can be used for this deformation tensor (strain tensor, mass tensor, etc.) but the classification can always follow the outline below:
\begin{itemize}
    \item 0 eigenvalue above the threshold: void,
    \item 1 eigenvalue above the threshold: wall,
    \item 2 eigenvalues above the threshold: filament,
    \item 3 eigenvalues above the threshold: knot.
\end{itemize}

Notably, the T-web classification developed by \cite{Forero_Romero_2009} is based on the tidal shear tensor $T$, which is the tensor of the second derivative of the gravitational potential
\begin{equation}
    T_{ij}=\frac{\partial^2\Phi}{\partial r_i r_j},
    \label{eq:Tweb}
\end{equation}
where $\Phi$ is the gravitational potential and $r_i$ with $i=1,2,3$ are the spatial coordinates. 
Alternatively, the V-web classification developed by \cite{Hoffman_2012} is based on the velocity shear tensor $\Sigma$:
\begin{equation}
 \Sigma_{ij} = -\frac{1}{2}\left(\frac{\partial v_{i}}{\partial r_j} + \frac{\partial v_{j}}{\partial r_{i}}\right) / H_0,
\end{equation}
where  $H_0$ is the Hubble constant and $v$ the velocity. We note that both of the above classifications are similar to the one used by \cite{Hahn_2007a}. Given those definitions, in practice, one can start from field maps, compute the deformation tensor, diagonalise it, and obtain the number of eigenvalues above a given threshold $\Lambda_\text{th}$ at each spatial position, thus dividing the cosmic web into four characteristic classes. 

Usually, the value of this threshold is not fixed a priori in the literature and is often chosen in order to obtain satisfactory visual agreement between the log density field and the classification. Originally, \cite{Hahn_2007a} took $\Lambda_\text{th}=0$ to distinguish structures with inner or outer flows. Follow-up studies demonstrated that having a non-zero threshold leads to a better visual match \citep[for example,][]{Forero_Romero_2009, Libeskind_2017, Suarez-Perez_2021, Cui_2017, Hoffman_2012, Carlesi_2014}. The threshold is generally taken as a positive constant of between $0$ and $1$, and usually does not depend on the redshift or smoothing scale, but we note that a positive threshold will tend to underline the most extreme knots and filaments (which is usually what is meant by good visual agreement) while a negative threshold will then underline the most empty voids. 

\begin{figure}
    \centering
    \includegraphics[width=\textwidth]{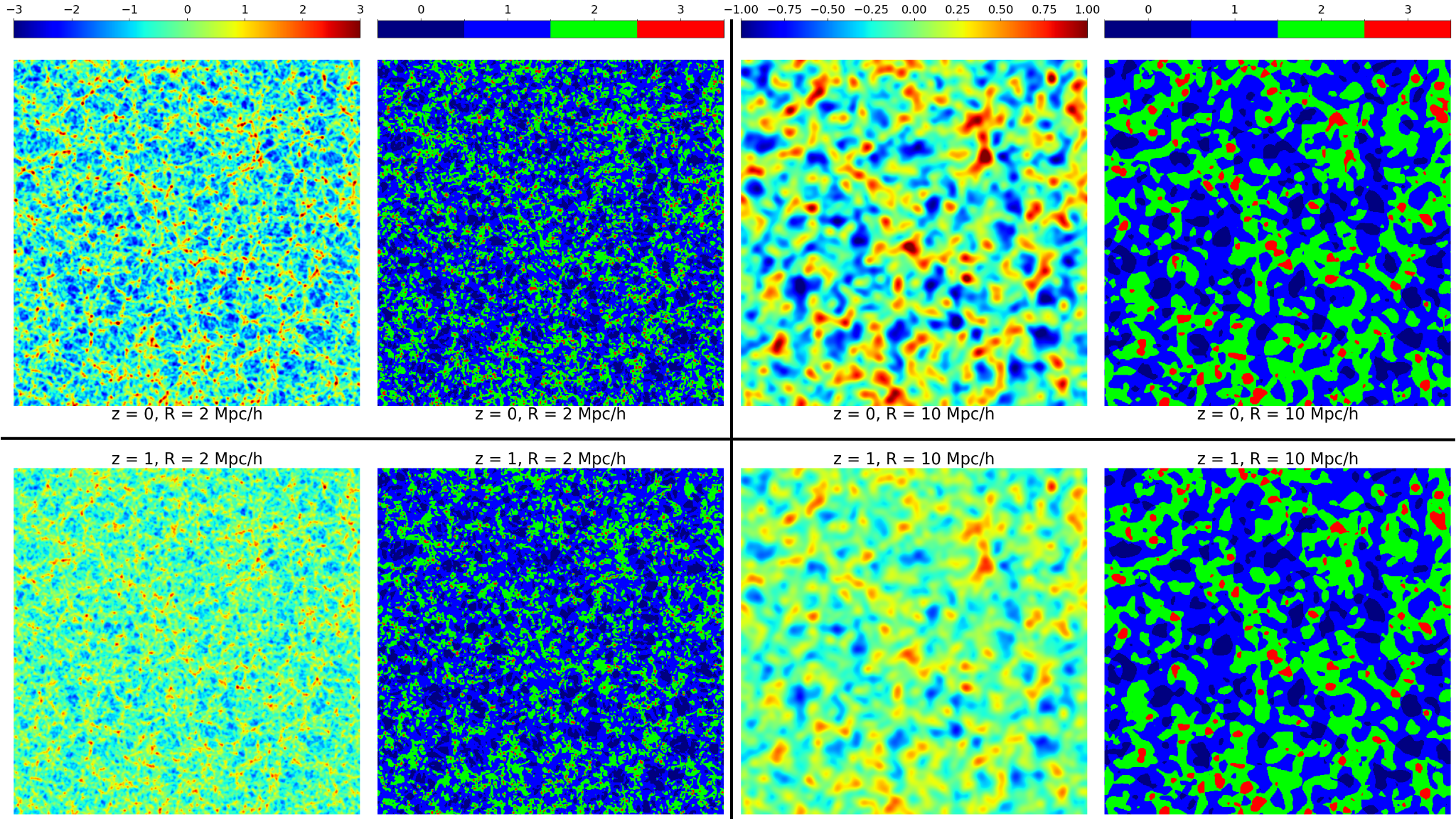}
    \caption{Logarithm of the density contrast of one slice of $1$ Gpc/h in height and width and $2$ Mpc/h in thickness of the Quijote simulation (first and third column) compared with the classification obtained using the T-web (second and fourth column) at different redshifts and smoothing scales. For the T-web classification, the colour bar is the number of eigenvalues above our chosen threshold $\Lambda_\text{th}$ = 0.01 chosen specifically for good visual agreement.}
    \label{fig:tranche}
\end{figure}

In the remainder of this paper, we focus on the T-web formalism, although we emphasise that the V-web description would yield identical results at first perturbative order in a Lagrangian framework (the so-called 1LPT or Zeldovitch approximation). This is indeed due to the usual expression for these ballistic trajectories where the velocity is given by the gradient of the gravitational potential up to a uniform time-dependent factor \citep{Zeldovich1970}. As an example of the accuracy of this classification scheme, Fig.~\ref{fig:tranche} presents a comparison of the density contrast (first and third columns, with a continuous colour bar) at different redshifts and smoothing scales, with the classification obtained using the T-web (second and fourth column, with a discrete color bar) at the same redshifts and smoothing scales. In the second and fourth columns, voids are shown in dark blue, walls in blue, filaments in green, and nodes in red. As expected from the T-web classification, we obtain the environments of the simulation by computing the second derivative of the potential and looking at the number of eigenvalues above our chosen threshold. Here, we are using a threshold of $\Lambda_{\text th} = 0.01$ for every redshift and smoothing scale based on the value used in \cite{Cui_2017}, which was chosen in order to have good visual agreement. For instance, the knots and voids (respectively in red and dark blue) can easily be identified by eye in both visualisations and are at the same position as rare maxima and minima in the density contrast maps. 

To build a theoretical description of the cosmic web as defined by the T-web description, let us first define the variance of the contrast of the density field, $\delta = (\rho - \langle \rho \rangle)/\langle \rho \rangle$, 
\begin{equation}
   \sigma_0^2 = \langle \delta^2\rangle . 
\end{equation}
Following \cite{pogosyan2009invariantGC}, we choose to normalise the derivatives of the gravitational potential by their variance, such that 
\begin{equation}
   \nu = \frac{1}{\sigma_0}\delta, \qquad \phi_{ij} = \frac{1}{\sigma_0}\nabla_i \nabla_j \Phi.
   \label{eq:norm_delta}
\end{equation}
Formally, given our classification method, the probability of each cosmic environment then depends on the joint probability distribution ${\cal P}$ of the eigenvalues of the tidal tensor. Given our normalisation choice in Eqs.~\eqref{eq:norm_delta}, here we use the eigenvalues normalised by their variance (or equivalently the eigenvalues of the normalised tidal tensor), which we denote $\lambda_1\leq\lambda_2\leq\lambda_3$.
The probabilities of the four environments can then be written as
\begin{align}
    &P_\text{void}=\int {\rm d}\lambda_1 {\rm d}\lambda_2 {\rm d}\lambda_3\;{\cal P}(\lambda_1,\lambda_2,\lambda_3) Boole(\lambda_1 < \lambda_2< \lambda_3 < \lambda_{\rm th})\label{eq:void} \\
    &P_\text{wall}=\int {\rm d}\lambda_1 {\rm d}\lambda_2 {\rm d}\lambda_3\;{\cal P}(\lambda_1,\lambda_2,\lambda_3) Boole(\lambda_1 < \lambda_2 < \lambda_{\rm th}< \lambda_3), \\
    &P_\text{filament}=\int {\rm d}\lambda_1 {\rm d}\lambda_2 {\rm d}\lambda_3\;{\cal P}(\lambda_1,\lambda_2,\lambda_3) Boole(\lambda_1< \lambda_{\rm th}< \lambda_2< \lambda_3), \\ 
    &P_\text{knot}=\int {\rm d}\lambda_1 {\rm d}\lambda_2 {\rm d}\lambda_3\;{\cal P}(\lambda_1,\lambda_2,\lambda_3)Boole(\lambda_{\rm th} < \lambda_1< \lambda_2< \lambda_3),\label{eq:knot}
\end{align}
where $\{\lambda_i\}_{i=1,2,3}$ are the orderly eigenvalues of the $(\phi_{ij})_{1\leq i,j\leq3}$ matrix and $Boole$ is a Boolean equal to 1 if the condition is satisfied, and 0 otherwise. 

As we are now working with normalised eigenvalues, and keeping our earlier choice of a $\Lambda_{\rm th} = 0.01$ value, our threshold will be given by $\lambda_{\rm th}={(\Lambda_{\rm th} = 0.01)}/{\sigma(z)}$, with
\begin{equation}
    \sigma^2(z) =  4 \pi \int {\rm d}k \, k^2 P(k,z)W(kR)^2,
    \label{eq:sig0}
\end{equation}
where $W(kR)$ is the applied smoothing, which in this paper is Gaussian with a smoothing scale $R$ such that
\begin{equation}
    W_\text{G}(kR) = \exp\left(-\frac{1}{2}k^2R^2\right),
\end{equation}
and $P(k,z)$ is the matter density power spectrum. We denote the linear variance $\sigma_0^2$ when using the linear power spectrum, which is computed using the Boltzmann code {\sc Camb} \citep{Lewis_1999}, and $\sigma_\text{NL}^2$ the non-linear variance (e.g. measured in the simulation or predicted with emulators for example).

\section{Cosmic web abundances in the linear regime} 
\label{gaussian}

To obtain some theoretical information about the abundance of the different cosmic environments in the density field and their redshift evolution, let us first consider the density field at linear order in Eulerian perturbation theory, this is, let us assume that it is Gaussian. For a Gaussian density field, the associated gravitational potential and its successive derivatives are also Gaussian-distributed, which means that we can use the Doroshkevich formula \citep{doroskevich1970Ap......6..320D}, which gives the joint probability distribution of eigenvalues of a Gaussian symmetric matrix:
\begin{equation} \label{eq:doro}
  {\cal P}_D(\lambda_1,\lambda_2,\lambda_3) =   \frac{675\sqrt{5} e^{\frac{3}{4}(\lambda_1 + \lambda_2 + \lambda_3)^2 - \frac{15}{4}(\lambda_1^2 + \lambda_2^3 + \lambda_3^2)}}{8 \pi} (\lambda_3 - \lambda_2)(\lambda_2 - \lambda_1)(\lambda_3 - \lambda_1).
\end{equation}
We note that the general expression for the multi-dimensional PDF of the tidal tensor in Cartesian coordinates $X=\{\phi_{ij}\}$ in the  Gaussian case reads
\begin{equation} \label{eq:gaussienne}
    {\cal P_G}(X) = (2 \pi)^{-N/2}|C|^{-1/2}\exp\left(-\frac{1}{2}XC^{-1}X\right),
\end{equation}
where $C$ is its covariance matrix.

From Eq.~\eqref{eq:doro}, we can then numerically integrate Eqs.~\eqref{eq:void}-\eqref{eq:knot} to obtain the probabilities of the various environments. In practice, those 3D integrals can be reduced to 1D as two degrees of freedom can be analytically integrated out. For that purpose, one needs to express the three degrees of freedom of the tidal tensor that are rotationally invariant as polynomials of its Cartesian coordinates $\phi_{ij}$ (in contrast to eigenvalues, which are not rotationally invariant). Following for instance \cite{pogosyan2009invariantGC}, a natural option can be the three coefficients of its characteristic polynomial:

\begin{eqnarray}
     I_1 &=& Tr(\phi_{ij}) = \phi_{11}+\phi_{22}+\phi_{33}  = \lambda_1+\lambda_2+\lambda_3 = \nu, \\
     I_2 &=& \phi_{11}\phi_{22} + \phi_{22}\phi_{33}+\phi_{11}\phi_{33}-\phi_{12}^2-\phi_{23}^2-\phi_{13}^2 = \lambda_1\lambda_2+\lambda_2\lambda_3+\lambda_1\lambda_3,\\
     I_3 &=& det|\phi_{ij}| =\phi_{11}\phi_{22}\phi_{33}+2\phi_{12}\phi_{23}\phi_{13}-\phi_{11}\phi_{23}^2-\phi_{22}\phi_{13}^2-\phi_{33}\phi_{12}^2 = \lambda_1\lambda_2\lambda_3.
\end{eqnarray}

To get variables that are as uncorrelated as possible, a simplification proposed by \cite{pogosyan2009invariantGC} is to use combinations of the $\{I_k\}_{1\geq k \geq 3}$ ---denoted $\{J_k\}_{1\geq k \geq 3}$--- such that the first variable is again the trace of the tidal tensor but higher-order variables only depend on the traceless part of that tensor and are thus uncorrelated with the trace in the Gaussian case. 
After some algebra, one can easily find
\begin{equation}
    J_1 = I_1, \qquad
    J_2 = I_1^2 - 3I_2, \qquad
    J_3 = I_1^3 -\frac{9}{2}I_1I_2 + \frac{27}{2}I_3.
\end{equation}
For a Gaussian random field, one can easily show that the probability distribution function of the above-defined variables reads \citep{pogosyan2009invariantGC} 

\begin{equation} \label{eq:gauss_gc}
    {\cal P_G}(J_1,J_2,J_3)=\frac{25\sqrt{5}}{12\pi}\exp\left[-\frac{1}{2}J_1^2-\frac{5}{2}J_2\right].
\end{equation}
where $J_2\geq 0$ and $J_3$ is uniformly distributed between its boundaries such that $-J_2^{3/2}\leq J_3\leq{J_2^{3/2}}$. Let us note that the probability distribution of Js can be easily mapped to the Doroskevitch formula for the distribution of the eigenvalues after taking into account the usual Vandermonde determinant.

To get the abundances for the different environments, we use a criterion on the number of eigenvalues superior to the threshold. As we are now working with variables $(J_1,J_2,J_3)$, we have to translate the criterion on eigenvalues into this new set of variables. Following \cite{pogosyan2009invariantGC} and \cite{Gay_2012GC}, we can obtain the integration limits for the four different environments as follows:

\begin{align}
    &P_\text{void}=\int_{0}^\infty {\rm d}J_2 \int_{-\infty}^{-2\sqrt{J_2}+3\lambda_{\rm th}}{\rm d}J_1 \int_{-J_2^{3/2}}^{J_2^{3/2}} {\rm d}J_3 \,{\cal P_G}(J_1,J_2,J_3) + \int_{0}^\infty {\rm d}J_2 \int_{-2\sqrt{J_2}+3\lambda_{\rm th}}^{-\sqrt{J_2}+3\lambda_{\rm th}} \!\!\!\!\!{\rm d}J_1 \int_{-J_2^{3/2}}^{-\frac{1}{2}(J_1-3\lambda_{\rm th})^3+\frac{3}{2}(J_1-3\lambda_{\rm th})J_2} \!\!\!\!\!\!\!\!\!\!\!\!\!\!\!\!\!\!\!\!\!\!\!\!\!\!\!\!\!\!\!\!\!\!\!\!\!\!\!\!\!\! {\rm d}J_3 {\cal P_G}(J_1,J_2,J_3), \label{eq:pv} \\
    &P_\text{wall}=\int_{0}^\infty {\rm d}J_2 \int_{-2\sqrt{J_2}+3\lambda_{\rm th}}^{\sqrt{J_2}+3\lambda_{\rm th}} {\rm d}J_1 \int_{-\frac{1}{2}(J_1-3\lambda_{\rm th})^3+\frac{3}{2}(J_1-3\lambda_{\rm th})J_2}^{J_2^{3/2}} \!\!\!\!\!\!\!\!\!\!\!\!\!\!\!\!\!\!\!\!\!\!\!\!\!\!\!\!\!\!\!\!\!\!\!\!\! {\rm d}J_3 {\cal P_G}(J_1,J_2,J_3), \label{eq:pw} \\
    &P_\text{filament}=\int_{0}^\infty {\rm d}J_2 \int_{-\sqrt{J_2}+3\lambda_{\rm th}}^{2\sqrt{J_2}+3\lambda_{\rm th}} {\rm d}J_1 \int_{-J_2^{3/2}}^{-\frac{1}{2}(J_1-3\lambda_{\rm th})^3+\frac{3}{2}(J_1-3\lambda_{\rm th})J_2} \!\!\!\!\!\!\!\!\!\!\!\!\!\!\!\!\!\!\!\!\!\!\!\!\!\!\!\!\!\!\!\!\!\!\!\!\! {\rm d}J_3 {\cal P_G}(J_1,J_2,J_3), \label{eq:pf} \\
    &P_\text{knot}=\int_{0}^\infty {\rm d}J_2 \int_{\sqrt{J_2}+3\lambda_{\rm th}}^{2\sqrt{J_2}+3\lambda_{\rm th}} {\rm d}J_1 \int_{-\frac{1}{2}(J_1-3\lambda_{\rm th})^3+\frac{3}{2}(J_1-3\lambda_{\rm th})J_2}^{J_2^{3/2}} \!\!\!\!\!\!\!\!\!\!\!\!\!\!\!\!\!\!\!\!\!\!\!\!\!\!\!\!\!\!\!\!\!\!\!\!\! {\rm d}J_3 {\cal P_G}(J_1,J_2,J_3) +\int_{0}^\infty {\rm d}J_2 \int_{2\sqrt{J_2}+3\lambda_{\rm th}}^{\infty} {\rm d}J_1 \int_{-J_2^{3/2}}^{J_2^{3/2}} {\rm d}J_3 {\cal P_G}(J_1,J_2,J_3). \label{eq:pk}
\end{align}
The analytical integration over $J_1$ and $J_3$ can now be performed and we thus obtain the following probabilities of the four environments (written in terms of 1D integrals over $J_2$):
\begin{align} \label{eq:pg_void}
\tiny
\begin{split}
    P_\text{void}&=\int_0^\infty \frac{25 \sqrt{5}}{48\pi} e^{-\frac{5 J_2}{2}}\Biggl[-\sqrt{2 \pi } \bigl(2 J_2^{3/2}-9 J_2 \text{$\lambda_{\rm th} $} + 9 \left(3 \text{$\lambda_{\rm th}$}^3+\text{$\lambda_{\rm th}$}\right)\bigr) \text{erf}\left(\frac{3 \text{$\lambda_{\rm th}$}-2 \sqrt{J_2}}{\sqrt{2}}\right)
    +\sqrt{2 \pi } \left(2 J_2^{3/2}-9 J_2 \text{$\lambda_{\rm th}$}+9 \left(3 \text{$\lambda_{\rm th}$}^3+\text{$\lambda_{\rm th}$}\right)\right) \text{erf}\left(\frac{3 \text{$\lambda_{\rm th}$}-\sqrt{J_2}}{\sqrt{2}}\right) \\
    &+ 4 \sqrt{2 \pi } J_2^{3/2} \text{erfc}\left(\frac{2 \sqrt{J_2}-3 \text{$\lambda_{\rm th}$}}{\sqrt{2}}\right)
    -2 e^{-\frac{1}{2} \left(2 \sqrt{J_2}-3 \text{$\lambda_{\rm th}$}\right)^2} \left(6 \sqrt{J_2} \text{$\lambda_{\rm th}$}+J_2+9 \text{$\lambda_{\rm th}$}^2+2\right) + e^{-\frac{1}{2} \left(\sqrt{J_2}-3 \text{$\lambda_{\rm th}$}\right)^2} \left(6 \sqrt{J_2} \text{$\lambda_{\rm th}$}-4 J_2+18 \text{$\lambda_{\rm th}$}^2+4\right)\Biggr]{\rm d}J_2,
\end{split}
\end{align}

\begin{align} \label{eq:pg_wall}
\tiny
\begin{split}
    P_\text{wall}&=\int_0^\infty \frac{25 \sqrt{5}}{48\pi} e^{-\frac{5 J_2}{2}} \Biggl[-\sqrt{2 \pi } (2 J_2^{3/2}+9 J_2 \text{$\lambda_{\rm th}$} - 9 \left(3 \text{$\lambda_{\rm th}$}^3+\text{$\lambda_{\rm th}$}\right)) \text{erf}\left(\frac{3 \text{$\lambda_{\rm th}$}-2 \sqrt{J_2}}{\sqrt{2}}\right)
    +\sqrt{2 \pi } \left(2 J_2^{3/2}+9 J_2 \text{$\lambda_{\rm th}$}-9 \left(3 \text{$\lambda_{\rm th}$}^3+\text{$\lambda_{\rm th}$}\right)\right) \text{erf}\left(\frac{\sqrt{J_2}+3 \text{$\lambda_{\rm th}$}}{\sqrt{2}}\right) \\
    &+ 2 e^{-\frac{1}{2} \left(2 \sqrt{J_2}-3 \text{$\lambda_{\rm th}$}\right)^2} \left(6 \sqrt{J_2} \text{$\lambda_{\rm th}$}+J_2+9 \text{$\lambda_{\rm th}$}^2+2\right)
    +2 e^{-\frac{1}{2} \left(\sqrt{J_2}+3 \text{$\lambda_{\rm th}$}\right)^2} \left(3 \sqrt{J_2} \text{$\lambda_{\rm th}$}+2 J_2-9 \text{$\lambda_{\rm th}$}^2-2\right)\Biggr]{\rm d}J_2,
\end{split}
\end{align}

\begin{align} \label{eq:pg_fila}
\tiny
\begin{split}
    P_\text{filament} &=\int_0^\infty -\frac{25 \sqrt{5}}{48\pi} e^{-\frac{5 J_2}{2}} \Biggl[-\sqrt{2 \pi } (2 J_2^{3/2}-9 J_2 \text{$\lambda_{\rm th}$} + 9 \left(3 \text{$\lambda_{\rm th}$}^3+\text{$\lambda_{\rm th}$}\right)) \text{erf}\left(\frac{2 \sqrt{J_2}+3 \text{$\lambda_{\rm th}$}}{\sqrt{2}}\right)
    +\sqrt{2 \pi } \left(2 J_2^{3/2}-9 J_2 \text{$\lambda_{\rm th}$}+9 \left(3 \text{$\lambda_{\rm th}$}^3+\text{$\lambda_{\rm th}$}\right)\right) \text{erf}\left(\frac{3 \text{$\lambda_{\rm th}$}-\sqrt{J_2}}{\sqrt{2}}\right) \\
    &- 2 e^{-\frac{1}{2} \left(2 \sqrt{J_2}+3 \text{$\lambda_{\rm th}$}\right)^2} \left(-6 \sqrt{J_2} \text{$\lambda_{\rm th}$}+J_2+9 \text{$\lambda_{\rm th}$}^2+2\right)
    +e^{-\frac{1}{2} \left(\sqrt{J_2}-3 \text{$\lambda_{\rm th}$}\right)^2} \left(6 \sqrt{J_2} \text{$\lambda_{\rm th}$}-4 J_2+18 \text{$\lambda_{\rm th}$}^2+4\right)\Biggr]{\rm d}J_2,
\end{split}
\end{align}

\begin{align} \label{eq:pg_knot}
\tiny
\begin{split}
    P_\text{knot}&=\int_0^\infty \frac{25 \sqrt{5}}{48 \pi } e^{-\frac{5 J_2}{2}} \Biggl[-\sqrt{2 \pi } (2 J_2^{3/2}+9 J_2 \text{$\lambda_{\rm th}$} - 9 \left(3 \text{$\lambda_{\rm th}$}^3+\text{$\lambda_{\rm th}$}\right)) \text{erf}\left(\frac{\sqrt{J_2}+3 \text{$\lambda_{\rm th}$}}{\sqrt{2}}\right)
    +\sqrt{2 \pi } \left(2 J_2^{3/2}+9 J_2 \text{$\lambda_{\rm th}$}-9 \left(3 \text{$\lambda_{\rm th}$}^3+\text{$\lambda_{\rm th}$}\right)\right) \text{erf}\left(\frac{2 \sqrt{J_2}+3 \text{$\lambda_{\rm th}$}}{\sqrt{2}}\right) \\
    &+ 4 \sqrt{2 \pi } J_2^{3/2} \text{erfc}\left(\frac{2 \sqrt{J_2}+3 \text{$\lambda_{\rm th}$}}{\sqrt{2}}\right)
    -2 e^{-\frac{1}{2} \left(2 \sqrt{J_2}+3 \text{$\lambda_{\rm th}$}\right)^2} \left(-6 \sqrt{J_2} \text{$\lambda_{\rm th}$}+J_2+9 \text{$\lambda_{\rm th}$}^2+2\right) + e^{-\frac{1}{2} \left(\sqrt{J_2}+3 \text{$\lambda_{\rm th}$}\right)^2} \left(-6 \sqrt{J_2} \text{$\lambda_{\rm th}$}-4 J_2+18 \text{$\lambda_{\rm th}$}^2+4\right)\Biggr]{\rm d}J_2,
\end{split}
\end{align}
where $\text{erf}(z)$ is the error function $\text{erf}(z) = {2}\int_0^z e^{-t^2}dt/{\sqrt{\pi}}$ and $\text{erfc}(z)$ is the complementary error function $\text{erfc}(z) = 1 - \text{erf}(z).$

\begin{figure}
    \centering
    \includegraphics[width=\textwidth]{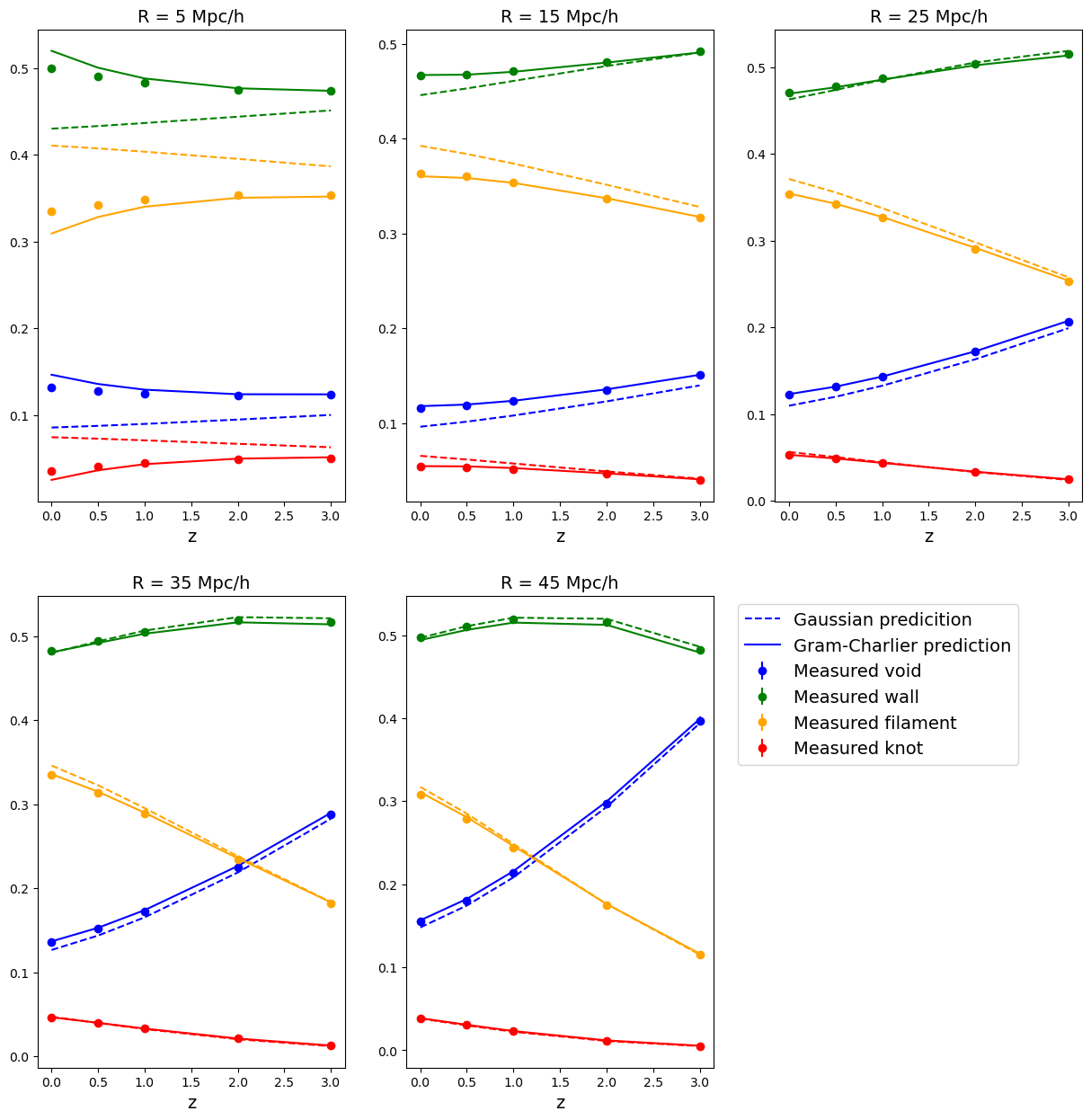}
    \caption{Probabilities of voids, walls, filaments, and knots as a function of the redshift. These probabilities are shown in blue, green, orange, and red, respectively. Each panel is obtained at a given smoothing scale. Dots are measurements from the simulation, dashed lines are the Gaussian prediction, and solid lines are the prediction obtained with the Gram-Charlier formalism at next-to-leading order. The error bars are errors on the mean but are too small to be distinguished.}
    \label{fig:proba}
\end{figure}

It is then possible to perform numerical integrations and obtain the abundance of each environment. This is illustrated in Fig.~\ref{fig:proba}, where the probabilities of the different environments are shown as a function of redshift. Each panel corresponds to a different Gaussian smoothing scale and displays the probability of voids, walls, filaments, and knots in blue, green, yellow, and red, respectively. The dots are the measurements obtained from the simulation, the dashed lines represent the Gaussian prediction obtained with the formalism described in this section, and the solid lines ---which can be ignored for now--- are the predictions at next-to-leading order obtained with a Gram-Charlier expansion, which is described in section~\ref{GC}, below. As expected, we observe that the higher the redshift and/or the larger the smoothing scale ---and thus the closer the simulation gets to the linear regime---, the closer the Gaussian prediction to the measurements. At lower redshift and smaller smoothing scales, non-Gaussianities are more important and departures from the Gaussian prediction thus appear. Here, all probabilities are computed using a threshold $\lambda_{\rm th}={0.01}/{\sigma_\text{NL}}$ inspired from the literature (using simulations) for which the evolution of $\sigma$ with redshift and smoothing scale is given in Table~\ref{tab:sig}. 
We note that the variance used solely in the normalisation of the threshold is the non-linear variance measured from the simulation. This allows us to have a meaningful threshold ---in terms of rarity--- even though we describe the cosmic structures in linear theory. Given the good agreement with the simulation already, that is, simply with Gaussian theory for sufficiently large scale and redshift, this states that the probabilities of the different environments are roughly captured by the statistics of a Gaussian field, at least for redshifts and scales that typically correspond to typical variances of $\sigma \lesssim 0.1$. 
Consequently, the redshift evolutions seen in previous works \citep{Cui_2017,Cui_2019} with a fixed threshold can be roughly understood simply as the non-linear evolution of the amplitude of fluctuations for that threshold. This is because a fixed threshold in non-linear densities does not correspond to a fixed 'rarity' or 'abundance' threshold for which the cosmic evolution would be much less important. This interpretation is notably valid at sufficiently large scale and redshift. Beyond $\sigma\sim 0.1$, such a Gaussian field approximation starts to break down. In the following section, we show how the accuracy of the theoretical model with the help of Gram-Charlier corrections.

\begin{table}[t]

  \caption{Standard deviation at different redshifts and (Gaussian) smoothing scales. The first value is the measurement from the Quijote simulation $\sigma_\text{NL}$, while the second column in parenthesis is the linear prediction $\sigma_0$.}
\label{tab:sig}
\begin{tabular}{lccccccc}
\hline
\hline
    z / R (Mpc/h) & $5 $ & $15 $ & $25 $ & $35 $ & $45 $ & $55 $ & $65 $ 
    \\ \hline
    $0$ & $0.71$ ($0.68$) & $0.26$ ($0.26$) & $0.15$ ($0.15$) & $0.10$ ($0.10$) & $0.074$ ($0.072$) & $0.057$ ($0.055$) & $0.045$ ($0.043$) \\ 
    $0.5$ & $0.54$ ($0.52$) & $0.20$ ($0.20$) & $0.12$ ($0.12$) & $0.078$ ($0.077$) & $0.57$ ($0.056$) & $0.044$ ($0.042$) & $0.34$ ($0.033$)  \\ 
    $1$ & $0.42$ ($0.41$) & $0.16$ ($0.16$) & $0.091$ ($0.091$) & $0.061$ ($0.061$) & $0.045$ ($0.044$) & $0.034$ ($0.033$) & $0.027$ ($0.026$) \\ 
    $2$ & $0.29$ ($0.28$) & $0.11$ ($0.11$) & $0.063$ ($0.062$) & $0.042$ ($0.042$) & $0.031$ ($0.030$) & $0.024$ ($0.023$) & $0.019$ ($0.018$)  \\ 
    $3$ & $0.21$ ($0.22$) & $0.082$ ($0.082$) & $0.047$ ($0.047$) & $0.032$ ($0.031$) & $0.023$ ($0.023$) & $0.018$ ($0.017$) & $0.014$ ($0.014$)  \\ \hline
  \end{tabular}

\end{table}

\begin{figure}
    \centering
    \includegraphics[width=\textwidth]{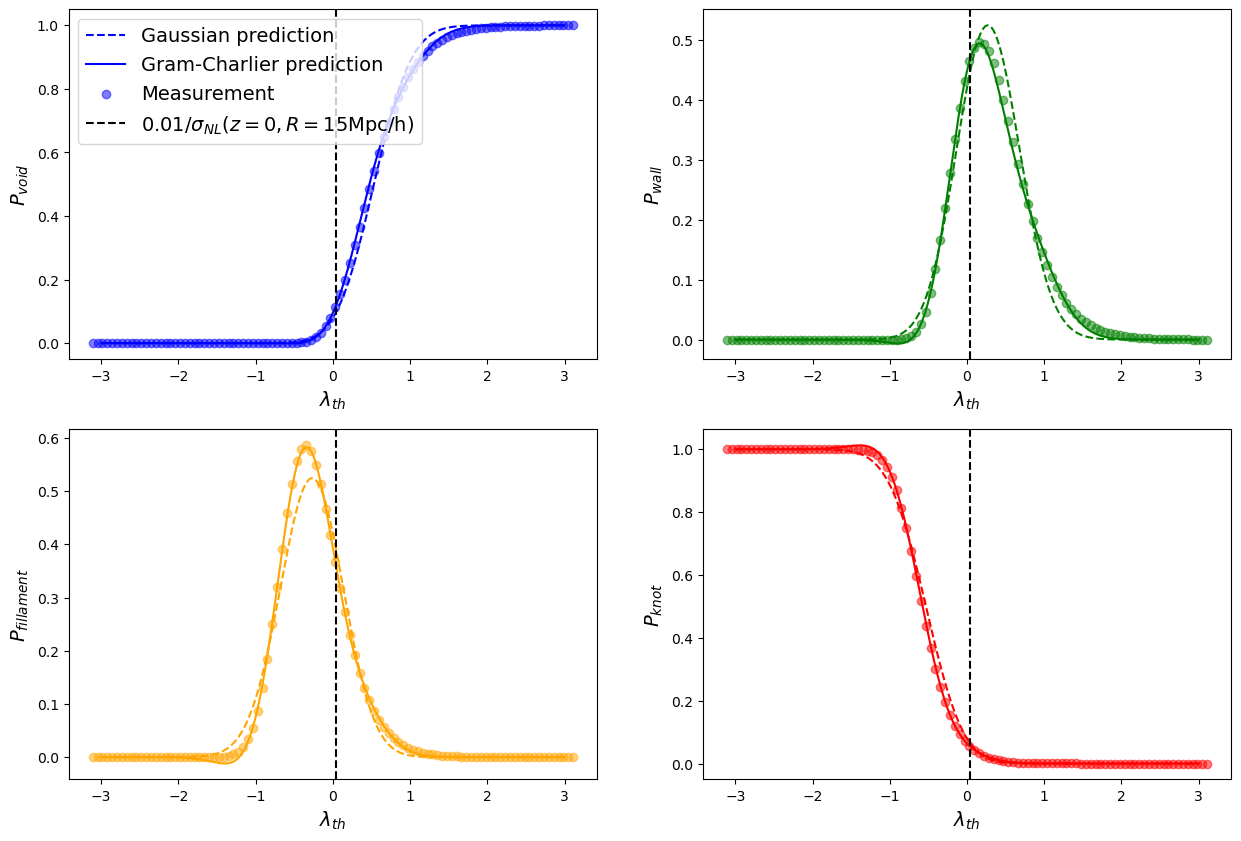}
    \caption{Probability of the environment as a function of the threshold for a Gaussian random field (dashed lines), Gram-Charlier (solid lines), and measurement in the simulations (dots) at R=15 Mpc/h and z=0. The vertical black line is the threshold $\Lambda_{\rm th}=0.01$ (thus $\lambda_\text{th} = \Lambda_{\rm th}/\sigma_\text{NL}$) used in this study. } 
    \label{fig:G_P}
\end{figure}

Before turning to the case of non-Gaussian corrections, let us emphasise that our Gaussian theoretical formalism allows us to obtain the probability of voids, walls, filaments, and knots as a function of the threshold itself, as illustrated in Fig.~\ref{fig:G_P} for redshift $z = 0$ and Gaussian smoothing R $= 15$ Mpc/h. The threshold used in the above analysis ($\Lambda_{\rm th}$ = 0.01) is the vertical black dashed line. For this choice of smoothing and redshift ---which corresponds to a mildly non-linear regime--- and for all the environments, we can draw the same conclusion as before: the qualitative picture is correctly captured but non-linear corrections are nonetheless necessary to improve the predictive power of our model.  This figure could be helpful to guide the choice of threshold based on theoretical arguments. We note that an alternative approach could be to renounce defining a global threshold and choose it according to the studied environment(s). For example, one could determine the threshold that gives the 20\% rarest knots and use the same threshold for the filaments, and obtain a threshold for the voids and walls in a similar fashion (which would be the opposite of the one used for knots and filaments in the simple Gaussian case). In this case, some spatial position may be in none of the environments and this will mean that this position is a transition between two environments. Another possibility would be to use the filling factor approach to fix a threshold in abundance (i.e we fix the volume fraction occupied by the excursion above $\nu$ \citep{Gott87,Matsubara}: this would lead to a remapping of our environments.

Hereafter, we adhere to the standard global-threshold strategy. The description and inclusion of the above-mentioned non-linear theoretical corrections are now the goals of the remainder of this paper.

\section{Mild non-Gaussian corrections to cosmic web abundances} \label{GC}
\label{part2}
Relying on Gaussian random fields is valid only in the linear regime of structure formation. However, at low redshift and small scales, increasing numbers of non-Gaussianities appear in the density field and corrections to the Gaussian predictions need to be accounted for. To improve our previous Gaussian predictions, we now propose to work with a probability distribution function ${\cal P}(X)$ that is no more Gaussian but includes non-linearities in a perturbative manner. In practice, we use a Gram-Charlier expansion following previous works in the literature including \cite{pogosyan2009invariantGC},\cite{Gay_2012GC} and \cite{2013MNRAS.435..531C}.

\subsection{Gram-Charlier expansion of the joint distribution}
\label{work2}
For a set $X$ of random fields, the Gram-Charlier expansion of the joint PDF reads
\begin{equation}
    {\cal P}(X) = {\cal P_G}(X) \left[ 1 + \sum_{n=3}^{\infty}\frac{1}{n!} Tr[\langle X^n\rangle _{GC}. h_n(X)] \right],
\end{equation}
where ${\cal P_G}(X)$ is a Gaussian kernel as defined in Eq.~\eqref{eq:gaussienne}, $h_n(x)$ are the Hermite tensors
$h_n(X) = (-1)^n{\cal P_G}^{-1}(X)\partial^n {\cal P_G}(X)/ \partial X^n$ and $\langle X^n \rangle _{GC} = \langle h_n(X) \rangle$ are the Gram Charlier coefficients. 

For our rotation-invariant variables $J_1,J_2,J_3$, this translates into the following expression
\begin{equation}
\begin{split}
    {\cal P}(J_1,J_2,J_3)&={\cal P_G}(J_1,J_2,J_3)\Biggl[ 1+ \sum_{n=3}^\infty \sum_{k,l}^{k+2l=n} \frac{(-1)^{l}5^l\times 3}{k!(3+2l)!!}\langle J_1^kJ_2^l \rangle_{GC} H_k(J_1)L_l^{(3/2)}\left( \frac{5}{2}J_2\right) \\
    &+\sum_{n=3}^\infty \sum_{k}^{k+3=n} \frac{25}{k!\times 21}\langle J_1^kJ_3\rangle_{GC} H_k(J_1)J_3 + \sum_{n=5}^\infty \sum_{k,l,m=1}^{k+2l+3m=n} \frac{c_{lm}}{k!}\langle J_1^kJ_2^lJ_3^m\rangle_{GC} H_k(J_1)F_{lm}(J_2,J_3)\Biggr],
\end{split}
\end{equation}
where ${\cal P_G}(J_1,J_2,J_3)$ is the Gaussian case given by Eq.~\eqref{eq:gauss_gc}, $H_n$ are the successive Hermite polynomials, $L_l^{(\alpha)}(x)$ are the generalised Laguerre polynomials, $c_{lm}$ are the normalisation coefficients that we leave undetermined and the orthogonal polynomials $F_{lm}$ associated with $J_2$ and $J_3$ are such that
\begin{equation*}
    \int_0^\infty {\rm d}J_2 \int_{-J_2^{3/2}}^{J_2^{3/2}}{\rm d}J_3 {\cal P_G}(J_1,J_2,J_3) F_{lm}(J_2,J_3)F_{l'm'}(J_2,J_3)=\delta_l^{l'}\delta_m^{m'} .
\end{equation*} 
We have, for example, these special cases: $F_{l0}=\sqrt{3\times 2^l \times l!/(3+2l)!!}\times L_l^{3/2}({5}J_2/2)$ and $F_{01}={5}J_3/{\sqrt{21}}$.

Here, we focus on the first corrective term, $n=3$, such that
\begin{equation}
\begin{split}
    {\cal P}(J_1,J_2,J_3)&={\cal P_G}(J_1,J_2,J_3)\Biggl[ 1+ \frac{1}{6}\langle J_1^3 \rangle_{GC} H_3(J_1)-\langle J_1 J_2 \rangle_{GC} H_1(J_1)L_1^{(3/2)}\left( \frac{5}{2}J_2\right)+\frac{25}{21}\langle J_3\rangle_{GC}J_3\Biggr] + o(\sigma_0^2).
\end{split}
\end{equation}
The Gram-Charlier cumulants read
\begin{equation}
    \langle J_1^3 \rangle_{GC} = \langle H_3(J_1)\rangle, \qquad
     \langle J_1J_2\rangle_{GC} = -\frac{2}{5}\left\langle H_1(J_1)L_1^{(3/2)}\left( \frac{5}{2}J_2 \right)\right\rangle, \qquad
    \langle J_3 \rangle_{GC} = \langle J_3 \rangle,
\end{equation}
such that we finally obtain the Gram-Charlier expression at first non-linear order (NLO):
\begin{equation}
\label{eq:gc}
    {\cal P}(J_1,J_2,J_3)={\cal P_G}(J_1,J_2,J_3)\Biggl[ 1+ \frac{1}{6} \langle J_1^3  \rangle (J_1^3-3J_1)+\frac{5}{2}\langle J_1J_2\rangle J_1 (J_2-1)+ \frac{25}{21}\langle J_3\rangle J_3\Biggr] + o(\sigma_0^2).
\end{equation}

\label{sec:cum}

Three cumulants appear in the next-to-leading-order Gram-Charlier probability distribution function (Eq.~\eqref{eq:gc}): $\langle J_1^3 \rangle$, $\langle J_1 J_2 \rangle$ and $\langle J_3 \rangle$. In Eulerian perturbation theory, those cumulants are linear in $\sigma_0$ at tree order \footnote{We remind the reader that we are working with fields renormalised by their variance, so that $\langle J_1^3 \rangle = \langle \delta^3 \rangle/\sigma_0^3$.} and we therefore introduce the reduced cumulants $S_3 = \langle J_1^3 \rangle / \sigma_0$, $U_3 = \langle J_1 J_2 \rangle / \sigma_0$ and $V_3 = \langle J_3 \rangle / \sigma_0$, which are constant in time at tree order. For example, $S_3$ is the usual cosmological skewness, whose analytical prediction at tree order is well known \citep[for example,][]{Peebles, Juszkiewicz_1993, 1995Lokas,Colombi_2000}. The other two reduced cumulants can be computed in a similar manner, described in Appendix~\ref{a:cum}. Fig.~\ref{fig:cumulants} shows the resulting tree-order cumulants as a function of the smoothing scale in dashed black, compared to the measurements in the simulation at different redshifts (as shown with different colours). We see a good agreement at almost all smoothing scales and redshifts as the prediction is almost always within the error bars. For large smoothing scales, the error bars increase due to the finite volume of the Quijote simulation which thus misses large wave modes. As expected, at low redshift and small smoothing scales, departures from tree-order predictions are seen as the non-linearities increase. In this work, we focus on mildly non-linear scales (about 10 Mpc/h and above) where a perturbative treatment is accurate, as illustrated by the values of the cumulants depicted in this figure.

\begin{figure}
    \centering
    \includegraphics[width=\textwidth]{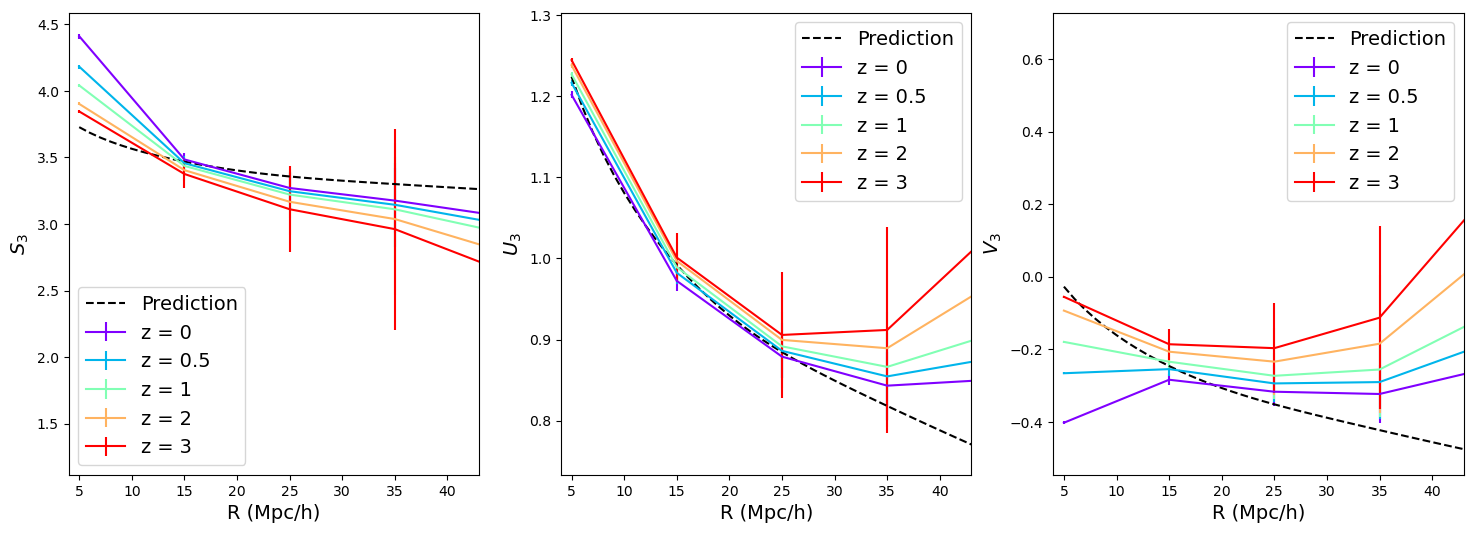}
    \caption{Cumulants $S_3$, $U_3$, and $V_3$ as a function of the smoothing scale R. The dashed black line is the analytical prediction. The coloured lines are the measurements from the simulations at different redshifts.}
    \label{fig:cumulants}
\end{figure}

\subsection{Cosmic web abundances at next-to-leading order} \label{res-proba}

We now turn to the explicit computation of the probability of different cosmic environments at next-to-leading order in the Gram-Charlier formalism. Let us first rewrite the Gram-Charlier expansion of the joint distribution of the rotational invariants of the tidal tensor as
\begin{equation}
    {\cal P}(J_1, J_2, J_3) = {\cal P_G}(J_1, J_2, J_3)+\sigma_0{\cal P}^{(3)}(J_1, J_2, J_3) + O(\sigma_0^2),
\end{equation}
where $\cal{P_G}$ is the Gaussian part and ${\cal P}^{(3)}$ is the first non-Gaussian corrective term
\begin{equation}
    {\cal P}^{(3)}(J_1,J_2,J_3)=\frac{25 \sqrt{5}}{12 \pi } \exp \left[-\frac{J_1^2}{2}-\frac{5 J_2}{2}\right]  \left(\frac{1}{6} \left(J_1^3-3 J_1\right) S_3+\frac{5}{2}J_1 (J_2-1) U_3+\frac{25}{21} J_3 V_3\right).
\end{equation}
Integrating this distribution function over the constraints given in Eqs.~\eqref{eq:pv} to \eqref{eq:pk} will give us the expression of the probability of a given environment $(E)$ as
\begin{equation} \label{eq:gc_env}
\begin{split}
    P_{(E)}(\lambda_{\rm th})&= P_{G,(E)}(\lambda_{\rm th})+\sigma_0 P^{(3)}_{(E)}(\lambda_{\rm th}) \\
    &= P_{G,(E)}(\lambda_{\rm th})+\sigma_0\bigl[s_{(E)}(\lambda_{\rm th})S_3+u_{(E)}(\lambda_{th,})U_3+v_{(E)}(\lambda_{\rm th})V_3\bigr],
\end{split}
\end{equation}
where $s,u$ and $v$ are threshold-dependent functions and where the Gaussian probability of the environments was derived in Eqs.~\eqref{eq:pg_void} to \eqref{eq:pg_knot}.

Again, two degrees of freedom can be integrated out and one remains to be integrated numerically. The resulting expressions for $s(\lambda_{\rm th})$, $u(\lambda_{\rm th})$, and $v(\lambda_{\rm th})$ in each environment are given below (written in terms of 1D integrals over $J_2$).

\subsubsection{Void}
For voids, we get
\begin{align}
\tiny
\begin{split}
    s_\text{void}(\lambda_{\rm th})&=\int_0^\infty \frac{25 \sqrt{5}}{144 \pi } e^{\frac{1}{2} (-9) \left(J_2+\lambda_{\rm th}^2\right)} \Biggl[-3 \sqrt{2 \pi } e^{2 J_2+\frac{9 \lambda_{\rm th}^2}{2}} \text{erf}\left(\frac{2 \sqrt{J_2}-3 \lambda_{\rm th}}{\sqrt{2}}\right)+3 \sqrt{2 \pi } e^{2 J_2+\frac{9 \lambda_{\rm th}^2}{2}} \text{erf}\left(\frac{\sqrt{\text{J2}}-3 \lambda_{\rm th}}{\sqrt{2}}\right)\\
    &+\sqrt{J_2} e^{3 \sqrt{J_2} \lambda_{\rm th}} \bigl[e^{3 \sqrt{J_2} \lambda_{\rm th}} (-48 J_2^{3/2} \lambda_{\rm th}+16 J_2^2+36 J_2 \lambda_{\rm th}^2 - 27 \sqrt{J_2} \lambda_{\rm th}\\
    &+14 J_2+12)-6 e^{\frac{3 J_2}{2}}\bigr]\Biggr]
    +\frac{25 \sqrt{5}}{36 \pi } J_2^{3/2} e^{6 \sqrt{J_2} \lambda_{\rm th}-\frac{9 J_2}{2}-\frac{9 \lambda_{\rm th}^2}{2}} \left(1-\left(2 \sqrt{J_2}-3 \lambda_{\rm th}\right)^2\right){\rm d}J_2,
\end{split}
\end{align}

\begin{align}
\tiny
\begin{split}
    u_\text{void}(\lambda_{\rm th})&=\int_0^\infty -\frac{125 \sqrt{5}}{96 \pi } (J_2-1) e^{\frac{1}{2} (-9) \left(J_2+\lambda_{\rm th}^2\right)} \Biggl[-3 \sqrt{2 \pi } e^{2 J_2+\frac{9 \lambda_{\rm th}^2}{2}} \left(J_2-9 \lambda_{\rm th}^2-1\right) \text{erf}\left(\frac{2 \sqrt{J_2}-3 \lambda_{\rm th}}{\sqrt{2}}\right)
    +3 \sqrt{2 \pi } e^{2 J_2+\frac{9 \lambda_{\rm th}^2}{2}} \left(J_2-9 \lambda_{\rm th}^2-1\right) \text{erf}\left(\frac{\sqrt{J_2}-3 \lambda_{\rm th}}{\sqrt{2}}\right) \\
    &- 2 e^{6 \sqrt{J_2} \lambda_{\rm th}} \left(4 J_2^{3/2}+6 \sqrt{J_2}+9 \lambda_{\rm th}\right) + 6 e^{3 \sqrt{J_2} \lambda_{\rm th}+\frac{3 J_2}{2}} \left(\sqrt{J_2}+3 \lambda_{\rm th}\right)\Biggr]
    -\frac{125 \sqrt{5}}{12 \pi } (J_2-1) J_2^{3/2} e^{6 \sqrt{J_2} \lambda_{\rm th}-\frac{9 J_2}{2}-\frac{9 \lambda_{\rm th}^2}{2}}{\rm d}J_2,
\end{split}
\end{align}

\begin{align}
\tiny
\begin{split}
    v_\text{void}(\lambda_{\rm th})&=\int_0^\infty \frac{625 \sqrt{5}}{4032 \pi } e^{-\frac{5 J_2}{2}} \Biggl[-\sqrt{2 \pi } \biggl[-4 J_2^3+9 J_2^2 + 243 (5-2 J_2) \lambda_{\rm th}^4+81 ((J_2-4) J_2+5) \lambda_{\rm th}^2-18 J_2+729 \lambda_{\rm th}^6+15\biggr] \\
    &\times \text{erf}\left(\frac{3 \lambda_{\rm th}-2 \sqrt{J_2}}{\sqrt{2}}\right)+\sqrt{2 \pi } \biggl[-4 J_2^3+9 J_2^2+243 (5-2 J_2) \lambda_{\rm th}^4 + 81 ((J_2-4) J_2+5) \lambda_{\rm th}^2-18 J_2+729 \lambda_{\rm th}^6+15\biggr] \text{erf}\left(\frac{3 \lambda_{\rm th}-\sqrt{J_2}}{\sqrt{2}}\right)\\
    &+2 e^{-\frac{1}{2} \left(\sqrt{J_2}-3 \lambda_{\rm th}\right)^2} \biggl[J_2^{3/2} \left(-\left(45 \lambda_{\rm th}^2+13\right)\right)+12 J_2^2 \lambda_{\rm th} + 4 J_2^{5/2}-9 J_2 \lambda_{\rm th} \left(15 \lambda_{\rm th}^2+7\right)+3 \sqrt{J_2} \left(27 \lambda_{\rm th}^4+36 \lambda_{\rm th}^2+5\right)\\
    &+9 \lambda_{\rm th} \left(27 \lambda_{\rm th}^4+42 \lambda_{\rm th}^2+11\right)\biggr]-2 e^{-\frac{1}{2} \left(2 \sqrt{J_2}-3 \lambda_{\rm th}\right)^2} \biggl[J_2^{3/2} \left(4-36 \lambda_{\rm th}^2\right)+3 J_2^2 \lambda_{\rm th}+2 J_2^{5/2}+J_2 \left(18 \lambda_{\rm th}-54 \lambda_{\rm th}^3\right)\\
    &+6 \sqrt{J_2} \left(27 \lambda_{\rm th}^4+36 \lambda_{\rm th}^2+5\right)+9 \lambda_{\rm th} \left(27 \lambda_{\rm th}^4+42 \lambda_{\rm th}^2+11\right)\biggr]\Biggr]{\rm d}J_2,
\end{split}
\end{align}
where $\text{erf}(z)$ is, again, the error function.
Once that analytical expression is obtained, we can perform the numerical integration over $J_2$. We display $s(\lambda_{\rm th})$, $u(\lambda_{\rm th})$, and $v(\lambda_{\rm th})$ as a function of the threshold in \ref{fig:suv}.

\subsubsection{Wall}
For walls, the same procedure gives
\begin{align}
\tiny
\begin{split}
   s_\text{wall}(\lambda_{\rm th})&=\int_0^\infty \frac{25 \sqrt{5}}{48 \pi } e^{-\frac{9 J_2}{2}} \Biggl[-\sqrt{2 \pi } e^{2 J_2} \text{erf}\left(\frac{3 \lambda_{\rm th}-2 \sqrt{J_2}}{\sqrt{2}}\right) + \sqrt{2 \pi } e^{2 J_2} \text{erf}\left(\frac{\sqrt{J_2}+3 \lambda_{\rm th}}{\sqrt{2}}\right)+\sqrt{J_2} \left(-e^{\frac{1}{2} (-3) \lambda_{\rm th} \left(2 \sqrt{J_2}+3 \lambda_{\rm th}\right)}\right)\\
   &\times \left(e^{9 \sqrt{J_2} \lambda_{\rm th}} \left(-9 \sqrt{J_2} \lambda_{\rm th}+6 J_2+4\right)+2 e^{\frac{3 J_2}{2}}\right)\Biggr] {\rm d}J_2,
\end{split}
\end{align}

\begin{align}
\tiny
\begin{split}
   u_\text{wall}(\lambda_{\rm th})&=\int_0^\infty \frac{125 \sqrt{5}}{32 \pi } e^{-\frac{9 J_2}{2}} (J_2-1) \Biggl[-\sqrt{2 \pi } e^{2 J_2} \left(J_2-9 \lambda_{\rm th}^2-1\right) \text{erf}\left(\frac{\sqrt{J_2}+3 \lambda_{\rm th}}{\sqrt{2}}\right)
   +\sqrt{2 \pi } e^{2 J_2} \left(J_2-9 \lambda_{\rm th}^2-1\right) \text{erf}\left(\frac{3 \lambda_{\rm th}-2 \sqrt{J_2}}{\sqrt{2}}\right) \\
   &- 2 e^{\frac{1}{2} (-3) \lambda_{\rm th} \left(2 \sqrt{J_2}+3 \lambda_{\rm th}\right)} \biggl[e^{9 \sqrt{J_2} \lambda_{\rm th}} \left(2 \sqrt{J_2}+3 \lambda_{\rm th}\right) + e^{\frac{3 J_2}{2}} \left(\sqrt{J_2}-3 \lambda_{\rm th}\right)\biggr]\Biggr] {\rm d}J_2,
\end{split}
\end{align}

\begin{align}
\tiny
\begin{split}
   v_\text{wall}(\lambda_{\rm th})&=\int_0^\infty -\frac{625 \sqrt{5}}{4032 \pi } e^{-\frac{5 J_2}{2}} \Biggl[-\sqrt{2 \pi }\biggl[-4 J_2^3+9 J_2^2 + 243 (5-2 J_2) \lambda_{\rm th}^4+81 ((J_2-4) J_2+5) \lambda_{\rm th}^2-18 J_2 + 729 \lambda_{\rm th}^6+15\biggl] \text{erf}\left(\frac{3 \lambda_{\rm th}-2 \sqrt{J_2}}{\sqrt{2}}\right)\\
   &+\sqrt{2 \pi } \biggl[-4 J_2^3+9 J_2^2+243 (5-2 J_2) \lambda_{\rm th}^4 + 81 ((J_2-4) J_2+5) \lambda_{\rm th}^2-18 J_2+729 \lambda_{\rm th}^6+15\biggr] \text{erf}\left(\frac{\sqrt{J_2}+3 \lambda_{\rm th}}{\sqrt{2}}\right)\\
   &-2 e^{-\frac{1}{2} \left(2 \sqrt{J_2}-3 \lambda_{\rm th}\right)^2} \biggl[J_2^{3/2} \left(4-36 \lambda_{\rm th}^2\right)+3 J_2^2 \lambda_{\rm th}+2 J_2^{5/2} + J_2 \left(18 \lambda_{\rm th}-54 \lambda_{\rm th}^3\right)+6 \sqrt{J_2} \left(27 \lambda_{\rm th}^4+36 \lambda_{\rm th}^2+5\right) + 9 \lambda_{\rm th} \left(27 \lambda_{\rm th}^4+42 \lambda_{\rm th}^2+11\right)\biggr]\\
   &+2 e^{-\frac{1}{2} \left(\sqrt{J_2}+3 \lambda_{\rm th}\right)^2} \biggl[J_2^{3/2} \left(45 \lambda_{\rm th}^2+13\right)+12 J_2^2 \lambda_{\rm th}-4 J_2^{5/2}-9 J_2 \lambda_{\rm th} \left(15 \lambda_{\rm th}^2+7\right) - 3 \sqrt{J_2} \left(27 \lambda_{\rm th}^4+36 \lambda_{\rm th}^2+5\right)+9 \lambda_{\rm th} \left(27 \lambda_{\rm th}^4+42 \lambda_{\rm th}^2+11\right)\biggr]\Biggr] {\rm d}J_2.
\end{split}
\end{align}

\subsubsection{Filament}
For filaments, we get
\begin{align}
\tiny
\begin{split}
   s_\text{filament}(\lambda_{\rm th})&=\int_0^\infty -\frac{25 \sqrt{5}}{144 \pi } e^{\frac{1}{2} (-3) \left(4 \sqrt{J_2} \lambda_{\rm th}+3 J_2+3 \lambda_{\rm th}^2\right)} \Biggl[3 \sqrt{2 \pi } e^{\frac{1}{2} \left(2 \sqrt{J_2}+3 \lambda_{\rm th}\right)^2} \text{erf}\left(\frac{2 \sqrt{J_2}+3 \lambda_{\rm th}}{\sqrt{2}}\right)
   +3 \sqrt{2 \pi } e^{\frac{1}{2} \left(2 \sqrt{J_2}+3 \lambda_{\rm th}\right)^2} \text{erf}\left(\frac{\sqrt{J_2}-3 \lambda_{\rm th}}{\sqrt{2}}\right) \\
   &- 3 \sqrt{J_2} \left(2 e^{9 \sqrt{J_2} \lambda_{\rm th}+\frac{3 J_2}{2}}+9 \sqrt{J_2} \lambda_{\rm th}+6 J_2+4\right)\Biggr] {\rm d}J_2,
\end{split}
\end{align}

\begin{align}
\tiny
\begin{split}
   u_\text{filament}(\lambda_{\rm th})&=\int_0^\infty -\frac{125 \sqrt{5}}{48 \pi } (J_2-1) \Biggl[ 3 \sqrt{\frac{\pi }{2}} e^{-\frac{5 J_2}{2}} \left(J_2-9 \lambda_{\rm th}^2-1\right) \text{erf}\left(\frac{3 \lambda_{\rm th}-\sqrt{J_2}}{\sqrt{2}}\right) - \frac{3}{2} e^{\frac{1}{2} (-3) \left(4 \sqrt{J_2} \lambda_{\rm th}+3 J_2+3 \lambda_{\rm th}^2\right)} \\
   & \times \Biggl[\sqrt{2 \pi } e^{\frac{1}{2} \left(2 \sqrt{J_2}+3 \lambda_{\rm th}\right)^2} \left(J_2-9 \lambda_{\rm th}^2-1\right) \text{erf}\left(\frac{2 \sqrt{J_2}+3 \lambda_{\rm th}}{\sqrt{2}}\right) + 6 \lambda_{\rm th} \left(e^{9 \sqrt{J_2} \lambda_{\rm th}+\frac{3 J_2}{2}}-1\right)+2 \sqrt{J_2} \left(e^{9 \sqrt{J_2} \lambda_{\rm th}+\frac{3 J_2}{2}}+2\right)\Biggr]\Biggr] {\rm d}J_2,
\end{split}
\end{align}

\begin{align}
\tiny
\begin{split}
   v_\text{filament}(\lambda_{\rm th})&=\int_0^\infty \frac{625 \sqrt{5}}{4032 \pi } e^{-\frac{5 J_2}{2}} \Biggl[-\sqrt{2 \pi } [-4 J_2^3+9 J_2^2+243 (5-2 J_2) \lambda_{\rm th}^4+81 ((J_2-4) J_2+5) \lambda_{\rm th}^2 - 18 J_2+729 \lambda_{\rm th}^6+15] \text{erf}\left(\frac{3 \lambda_{\rm th}-\sqrt{J_2}}{\sqrt{2}}\right)\\
   &+\sqrt{2 \pi } [-4 J_2^3+9 J_2^2+243 (5-2 J_2) \lambda_{\rm th}^4+81 ((J_2-4) J_2+5) \lambda_{\rm th}^2 - 18 J_2+729 \lambda_{\rm th}^6+15] \text{erf}\left(\frac{2 \sqrt{J_2}+3 \lambda_{\rm th}}{\sqrt{2}}\right)-2 e^{-\frac{1}{2} \left(\sqrt{J_2}-3 \lambda_{\rm th}\right)^2} \\
   &\times [J_2^{3/2} \left(-\left(45 \lambda_{\rm th}^2+13\right)\right)+12 J_2^2 \lambda_{\rm th}+4 J_2^{5/2} - 9 J_2 \lambda_{\rm th} \left(15 \lambda_{\rm th}^2+7\right)+3 \sqrt{J_2} \left(27 \lambda_{\rm th}^4+36 \lambda_{\rm th}^2+5\right) + 9 \lambda_{\rm th} \left(27 \lambda_{\rm th}^4+42 \lambda_{\rm th}^2+11\right)]\\
   &+e^{-\frac{1}{2} \left(2 \sqrt{J_2}+3 \lambda_{\rm th}\right)^2} [8 J_2^{3/2} \left(9 \lambda_{\rm th}^2-1\right)+6 J_2^2 \lambda_{\rm th}-4 J_2^{5/2}-36 J_2 \lambda_{\rm th} \left(3 \lambda_{\rm th}^2-1\right)-12 \sqrt{J_2} \left(27 \lambda_{\rm th}^4+36 \lambda_{\rm th}^2+5\right) + 18 \lambda_{\rm th} \left(27 \lambda_{\rm th}^4+42 \lambda_{\rm th}^2+11\right)]\Biggr] {\rm d}J_2.
\end{split}
\end{align}

\subsubsection{Knots}
Finally, for knots:
\begin{align}
\tiny
\begin{split}
   s_\text{knot}(\lambda_{\rm th})&=\int_0^\infty \frac{25 \sqrt{5}}{144 \pi } e^{\frac{1}{2} (-3) \left(4 \sqrt{J_2} \lambda_{\rm th}+3 J_2+3 \lambda_{\rm th}^2\right)} \Biggl[-3 \sqrt{2 \pi } e^{\frac{1}{2} \left(2 \sqrt{J_2}+3 \lambda_{\rm th}\right)^2} \text{erf}\left(\frac{\sqrt{J_2}+3 \lambda_{\rm th}}{\sqrt{2}}\right) + 3 \sqrt{2 \pi } e^{\frac{1}{2} \left(2 \sqrt{J_2}+3 \lambda_{\rm th}\right)^2} \text{erf}\left(\frac{2 \sqrt{J_2}+3 \lambda_{\rm th}}{\sqrt{2}}\right)\\
   &-\sqrt{J_2} \biggl[48 J_2^{3/2} \lambda_{\rm th}+16 J_2^2+36 J_2 \lambda_{\rm th}^2 - 6 e^{3 \sqrt{J_2} \lambda_{\rm th}+\frac{3 J_2}{2}}+27 \sqrt{J_2} \lambda_{\rm th}+14 J_2+12\biggr]\Biggr]
   +\frac{25 \sqrt{5}}{36 \pi } J_2^{3/2} e^{\frac{1}{2} (-3) \left(4 \sqrt{J_2} \lambda_{\rm th}+3 J_2+3 \lambda_{\rm th}^2\right)} \left(12 \sqrt{J_2} \lambda_{\rm th}+4 J_2+9 \lambda_{\rm th}^2-1\right) {\rm d}J_2,
\end{split}
\end{align}

\begin{align}
\tiny
\begin{split}
   u_\text{knot}(\lambda_{\rm th})&=\int_0^\infty \frac{125 \sqrt{5}}{96 \pi } (J_2-1) e^{\frac{1}{2} (-3) \left(4 \sqrt{J_2} \lambda_{\rm th}+3 J_2+3 \lambda_{\rm th}^2\right)} \Biggl[-3 \sqrt{2 \pi } e^{\frac{1}{2} \left(2 \sqrt{J_2}+3 \lambda_{\rm th}\right)^2} \left(J_2-9 \lambda_{\rm th}^2-1\right) \text{erf}\left(\frac{2 \sqrt{J_2}+3 \lambda_{\rm th}}{\sqrt{2}}\right)\\
   &+3 \sqrt{2 \pi } e^{\frac{1}{2} \left(2 \sqrt{J_2}+3 \lambda_{\rm th}\right)^2} \left(J_2-9 \lambda_{\rm th}^2-1\right) \text{erf}\left(\frac{\sqrt{J_2}+3 \lambda_{\rm th}}{\sqrt{2}}\right) - 8 J_2^{3/2}-18 \lambda_{\rm th} \left(e^{3 \sqrt{J_2} \lambda_{\rm th}+\frac{3 J_2}{2}}-1\right)+6 \sqrt{J_2} \left(e^{3 \sqrt{J_2} \lambda_{\rm th}+\frac{3 J_2}{2}}-2\right)\Biggr]\\
   &+\frac{125 \sqrt{5}}{12 \pi } (J_2-1) J_2^{3/2} e^{\frac{1}{2} (-3) \left(4 \sqrt{J_2} \lambda_{\rm th}+3 J_2+3 \lambda_{\rm th}^2\right)} {\rm d}J_2,
\end{split}
\end{align}

\begin{align}
\tiny
\begin{split}
   v_\text{knot}(\lambda_{\rm th})&=\int_0^\infty -\frac{625 \sqrt{5}}{4032 \pi } e^{-\frac{5 J_2}{2}} \Biggl[-\sqrt{2 \pi } \biggl[-4 J_2^3+9 J_2^2 + 243 (5-2 J_2) \lambda_{\rm th}^4+81 ((J_2-4) J_2+5) \lambda_{\rm th}^2-18 J_2 + 729 \lambda_{\rm th}^6+15\biggr] \text{erf}\left(\frac{\sqrt{J_2}+3 \lambda_{\rm th}}{\sqrt{2}}\right)\\
   &+\sqrt{2 \pi } \biggl[-4 J_2^3+9 J_2^2 + 243 (5-2 J_2) \lambda_{\rm th}^4+81 ((J_2-4) J_2+5) \lambda_{\rm th}^2-18 J_2 + 729 \lambda_{\rm th}^6+15\biggr] \text{erf}\left(\frac{2 \sqrt{J_2}+3 \lambda_{\rm th}}{\sqrt{2}}\right)
    +e^{-\frac{1}{2} \left(2 \sqrt{J_2}+3 \lambda_{\rm th}\right)^2} \biggl[8 J_2^{3/2} \left(9 \lambda_{\rm th}^2-1\right)\\
    &+6 J_2^2 \lambda_{\rm th}-4 J_2^{5/2}-36 J_2 \lambda_{\rm th} \left(3 \lambda_{\rm th}^2-1\right) - 12 \sqrt{J_2} \left(27 \lambda_{\rm th}^4+36 \lambda_{\rm th}^2+5\right)+18 \lambda_{\rm th} \left(27 \lambda_{\rm th}^4+42 \lambda_{\rm th}^2+11\right)\biggr]
   +2 e^{-\frac{1}{2} \left(\sqrt{J_2}+3 \lambda_{\rm th}\right)^2} \biggl[J_2^{3/2} \left(-\left(45 \lambda_{\rm th}^2+13\right)\right)-12 J_2^2 \lambda_{\rm th} \\
   &+ 4 J_2^{5/2}+9 J_2 \lambda_{\rm th} \left(15 \lambda_{\rm th}^2+7\right)+3 \sqrt{J_2} \left(27 \lambda_{\rm th}^4+36 \lambda_{\rm th}^2+5\right) 
   - 9 \lambda_{\rm th} \left(27 \lambda_{\rm th}^4+42 \lambda_{\rm th}^2+11\right)\biggr]\Biggr] {\rm d}J_2.
\end{split}
\end{align}

Plugging the previous expressions for the $s,u,v$ functions into Eq.~\eqref{eq:gc_env} finally allows us to compute the next-to-leading-order correction to the (threshold-dependent) probability of the different cosmic environments in the T-web classification.

As for the Gaussian case, we choose to measure a rarity threshold in the simulations, and therefore we plug in the non-linear variance measured in the simulation as the normalisation factor appearing in the threshold $\lambda_{\rm th}$. An alternative option to try and keep a meaningful threshold selection in terms of abundance or rarity could be to work with a filling factor, as mentioned at the end of Section~\ref{gaussian}.

The probabilities of the different environments as functions of redshift are displayed in Fig.~\ref{fig:proba}; these were already partially described in section~\ref{gaussian} for the Gaussian case. We now focus on the solid lines, which represent the Gram-Charlier correction described in this section. For every environment, redshift, and smoothing scale, the results obtained with the Gram-Charlier always improve the prediction over the Gaussian case. The agreement between the simulation and our non-linear model is almost perfect; only at the most non-linear scales (very low redshift and small smoothing scale) do the predictions start to become slightly different from the measurements, although the NLO correction always improves upon the Gaussian case. We can also look at the abundance of voids, walls, filaments, and knots as a function of the threshold in Fig.~\ref{fig:G_P}, which was also already partially described in section~\ref{gaussian}. We see that the Gram-Charlier correction not only improves the prediction over the Gaussian case but provides us with an extremely accurate prediction at all the relevant thresholds and for almost the whole spectrum of field values. 
Let us look at the contribution of each function that appears in our Gram-Charlier expansion: $s_{(E)}(\lambda_{\rm th})$, $u_{(E)}(\lambda_{\rm th})$, and $v_{(E)}(\lambda_{\rm th})$ in Eq.~\eqref{eq:gc_env} as shown in Fig.~\ref{fig:suv}. 
We note that $S_3$ is always larger than $U_3$ and $V_3$, implying that $s(\lambda_\text{th})$ is typically the dominant correction to the Gaussian case (and is actually related to the non-linear evolution of the threshold). This is the reason why our Gaussian model for the evolution of the cosmic web abundances already gives a fairly accurate prescription. The higher-order non-linear corrections are then modulated by the functions $s,u,v$ depending on the chosen threshold.

\section{Discussion and conclusion} \label{conclusion}
Above, using the T-web classification with a rarity threshold $\lambda_\text{th}={(\Lambda_{\rm th} = 0.01)}/ {\sigma_\text{NL}}$ (a commonly $\Lambda_{\rm th}$ value used in the literature), we describe a theoretical framework to accurately model the probabilities of voids, walls, filaments, and knots in the cosmic web. More precisely, as this computation requires knowledge of the joint probability distribution of the eigenvalues of the Hessian of the gravitational potential, we model an analogous object, which is the joint distribution between maximally decorrelated, rotational invariants that are polynomial in the Cartesian matrix components.

We first focused on the linear regime of structure formation where the density field is Gaussian (assuming Gaussian initial conditions) and where the distribution of the eigenvalues of the Hessian of the gravitational potential ---and then a Gaussian field--- are known. We find in that case that normalising the rarity threshold $\lambda_{\rm th}$ by the non-linear variance is surprisingly accurate with respect to measurements of the environment probabilities in the Quijote simulation in regimes where $\sigma\lesssim 0.1$ and captures most of the redshift and scale dependence (see Fig.~\ref{fig:proba}).

To probe the mildly non-linear regime, we then accounted for corrections to the Gaussian case in a perturbative manner relying on a Gram-Charlier expansion at first non-linear order (Eq.~\eqref{eq:gc}) for the joint distribution of our rotational invariants, and tree-order Eulerian perturbation theory to compute the Gram-Charlier coefficients. We showed that this correction to the Gaussian case increases the accuracy of predictions for the probability of cosmic environments at all considered redshifts and smoothing scales with threshold $\lambda_\text{th}={0.01} /{\sigma_\text{NL}}$ (see Fig.~\ref{fig:proba}), but also when varying the threshold at fixed redshift and smoothing scale (see Fig.~\ref{fig:G_P} for a reference case at $z=0$ and $R = 15$ Mpc/h).

In conclusion, all redshift- and scale-dependent evolution of T-web cosmic abundances previously observed in numerical simulations are shown to be predictable from first principles. In addition, having a precise theoretical model not only allows us to benchmark simulations but also to understand how these features depend on choices of threshold or background cosmological model for example. This information can be readily extracted from the predictions we provide here (without the need to run simulations or post-process them, etc.). This is notably important when one wants to use these cosmic web elements as a cosmological probe (to go typically beyond the power spectrum), which requires a cost-effective inference pipeline. We leave investigations along this line for future works. Let us also emphasise that we assume Gaussian initial conditions in this work, but primordial non-Gaussianities could easily be added to the formalism (following e.g. \cite{2013MNRAS.435..531C}), in which case a careful study could be performed of the extent to which cosmic web abundance depends on the physics of the primordial Universe.

Looking at Fig.~\ref{fig:G_P} for the direct result of the threshold-dependent modulation of the non-Gaussian cumulants appearing in the Gram-Charlier expansion of Fig.~\ref{fig:suv}, we observe that the different probabilities can fluctuate significantly when changing the threshold, as was pointed out for instance by \cite{Forero_Romero_2009}, who studied the impact of the choice of the threshold on the percolation and fragmentation of the cosmic web. With the present formalism, all this information can be described from first principles, which can be very useful in defining a physically motivated threshold rather than relying on visual inspection. Working with the abundance or rarity of an environment or with a filling factor could definitely help in that direction.

Another possible extension of this work could be to perform a multi-scale analysis in order to study the properties of halos depending on their environment. The T-web description developed here would provide the constrained environment (as often used in simulations) while the fields on a smaller scale would characterise the halo. However, let us note that the T-web classification is a rather rough description of the cosmic web (being sensitive not only to the mass distribution but also its larger-scale environment) and more sophisticated frameworks have already been developed in the literature. Interestingly though, most analyses studying environmental effects have shown that the precise definition of the environment has typically (and maybe surprisingly) little effect on the result \citep{Libeskind_2017} which is additional motivation for the development of theoretical predictions for even rudimentary cosmic web estimators such as that described here. In essence, the physical effects at play in this context are mostly tidal effects, which are naturally captured in the T-web description \citep{2021MNRAS.504.1694R}.

\begin{acknowledgements}
EA is partly supported by a PhD Joint Programme between the CNRS and the University of Arizona. 
AB's work is supported by the ORIGINS excellence cluster. SC acknowledges financial support from Fondation MERAC and by the SPHERES grant ANR-18-CE31-0009 of the French {\sl Agence Nationale de la Recherche}. This work has made use of the Infinity Cluster hosted by Institut d'Astrophysique de Paris. We thank Stephane Rouberol for running this cluster smoothly for us. 
\end{acknowledgements}

\bibliographystyle{aa} 
\bibliography{bibli.bib} 

\begin{appendix} 
\section{Computation of cumulants}\label{a:cum}

As discussed in section~\ref{sec:cum} and shown in Eq.~\eqref{eq:gc}, three cumulants are needed to predict cosmic web abundances at NLO. 
To obtain the value of these three cumulants, we use tree-order Eulerian perturbation theory. 

Given the scaling of the tree-order cumulants with the linear variance, and as we are working with normalised variables, there is no redshift dependence of the reduced cumulants, that is the normalised cumulants appearing in Eq.~\eqref{eq:gc} divided by the linear standard deviation $\sigma_0$ \citep{Bernardeau_2002}. In this Appendix, we propose to derive the required cumulants at tree-order starting from the well-known skewness, before extending this result to the other cumulants that are needed.

\subsection{$S_3 = \langle J_1^3 \rangle/\sigma_0$}

We note that $J_1=\nu=\delta/\sigma_0$, meaning that $\langle J_1^3 \rangle = \langle \delta^3 \rangle / \sigma_0^3 = S_3 \sigma_0$, where $S_3$ is the usual redshift-independent reduced skewness parameter characterising the asymmetry of the probability distribution function. Its computation is a well-established result of Eulerian perturbation theory \citep{Peebles,Juszkiewicz_1993} and we repeat its derivation here, on which we then base the extension to other cumulants.

The core hypothesis of the Eulerian perturbation scheme is to assume that the density field can be expanded at every order as a function of the initial (linear) density so that it can be written as a series of the form $\delta({\bm x},t) = \sum_n \delta^{(n)}({\bm x},t)$. At leading order, we thus obtain $\langle \delta^3 \rangle \approx 3 \langle (\delta^{(1)})^2\delta^{(2)} \rangle$ and where
\begin{equation}
    \delta^{(2)}({\bm x},t) = \int \frac{{\rm d}^3 \bm{k_1}}{(2 \pi)^{3/2}} \frac{{\rm d}^3 \bm{k_2}}{(2 \pi)^{3/2}} \, \delta^{(1)}(\bm{k_2},t) \, \delta^{(1)}(\bm{k_2},t) \, F_2(\bm{k_1},\bm{k_2}) \, e^{i({\bm k_1} + {\bm k_2}) \cdot {\bm x}}
\end{equation}
with
\begin{equation}
    F_2(\bm{k_1},\bm{k_2}) = \frac{5}{7} + \frac{1}{2}\frac{\bm{k_1} \cdot \bm{k_2}}{k_1 k_2}(\frac{k_1}{k_2} + \frac{k_2}{k_1}) + \frac{2}{7}\frac{(\bm{k_1} \cdot \bm{k_2})^2}{k_1^2 k_2^2}.
\end{equation}

The computation of $\langle \delta^3 \rangle$ therefore leads to the appearance of ensemble averages of the form $\langle \delta^{(1)}(\bm{k_1}) \, \delta^{(1)}(\bm{k_2}) \, \delta^{(1)}(\bm{k_3}) \, \delta^{(1)}(\bm{k_4}) \rangle$, which for Gaussian initial conditions can be estimated using Wick's theorem only considering pairs of wave vectors. Taking into account a smoothing window function $W(kR)$ and after some algebra, we obtain
\begin{equation}
     \langle \delta^3 \rangle = 6 \int {\rm d}^3 \bm{k_1} \int {\rm d}^3 \bm{k_2} P(k_1)P(k_2)F_2(\bm{k_1},\bm{k_2}) W(k_1R)W(k_2R)W(|\bm{k_1+k_2}|R),
     \label{skewness}
\end{equation}
with $P(k)$ being the linear power spectrum.

In this paper, we chose to work with Gaussian smoothing so that
\begin{equation}
    W(kR) = \exp \left( -\frac{1}{2}k^2R^2\right),
\end{equation}

\begin{equation}
    W(|\bm{k_1+k_2}|R) = \exp\left(-\frac{1}{2}(k_1^2+k_2^2)R^2\right)\sum_{l=0}^\infty (-1)^l(2l+1)P_l\left( \frac{\bm{k_1.k_2}}{k_1k_2} \right)I_{l+1/2}(k_1k_2R^2)\sqrt{\frac{\pi}{2k_1k_2R^2}},
\end{equation}
and $R$ is our smoothing scale, $P_l$ are the Legendre polynomials, and $I_\alpha$ are the modified Bessel functions of the first kind. The angular integration of~\eqref{skewness} is then performed, also decomposing $F_2$ on the basis of Legendre polynomials,
\begin{equation}
    F_2=\frac{17}{21}P_0 + \frac{1}{2}\left(\frac{k_1}{k_2}+\frac{k_2}{k_1}\right)P_1 + \frac{4}{21}P_2,
\end{equation}
and using the Legendre polynomial orthogonality relation,
\begin{equation}
    \int_{-1}^{1}P_m(x)P_n(x){\rm d}x=\frac{2}{2m+1}\delta^{\rm Dirac}_{m,n}.
\end{equation}

Noting that the angular part of~\eqref{skewness} only depends on the angle between ${\bm k_1}$ and ${\bm k_2}$, we obtain
\begin{equation}
\begin{split}
    \langle \delta^3 \rangle  &= \int \frac{24}{\sqrt{R}} \sqrt{2} \pi ^{5/2} k_1^{3/2} k_2^{3/2} P(k_1) P(k_2) e^{R \left(-k_1^2-k_2^2\right)}  \\
    &\times\left(\frac{4 \sqrt{\frac{2}{\pi }} \left(\left(\frac{6}{k_1^2 k_2^2 R^2}+2\right) \sinh (k_1 k_2 R)-\frac{6 \cosh (k_1 k_2 R)}{k_1 k_2 R}\right)}{21 \sqrt{k_1 k_2 R}}+\frac{34 \sqrt{\frac{2}{\pi }} \sinh (k_1 k_2 R)}{21 \sqrt{k_1 k_2 R}}-\frac{\left(\frac{k_1}{k_2}+\frac{k_2}{k_1}\right) \left(2 \cosh (k_1 k_2 R)-\frac{2 \sinh (k_1 k_2 R)}{k_1 k_2 R}\right)}{\sqrt{2 \pi } \sqrt{k_1 k_2 R}}\right) {\rm d}k_1 {\rm d}k_2\,.
    \label{skweness_final}
\end{split}
\end{equation}

The final 2D integrations over $k_1$ and $k_2$ are then usually obtained numerically for a general power spectrum. However, this step can be performed analytically in the special case of a power-law linear power spectrum $P(k)\propto k^n$, where $n$ is the spectral index. In such a case, it yields
\begin{equation}
\begin{split}
    \langle \delta^3\rangle/\sigma_0^4 &= -\frac{1}{14n^2\Gamma^2(\frac{3+n}{2})} 3\Gamma^2\left(\frac{1+n}{2}\right) \Biggl[3(16+7n)_2F_1\left(\frac{1+n}{2},\frac{1+n}{2},-\frac{1}{2},\frac{1}{4}\right)\\
    &+\left(-32+n(2+n)(23+28n)\right)_2F_1\left(\frac{1+n}{2},\frac{1+n}{2},\frac{1}{2},\frac{1}{4}\right)-7n(1+n)(3+2n)_2F_1\left(\frac{1+n}{2},\frac{3+n}{2},\frac{1}{2},\frac{1}{4}\right)\Biggl],
\end{split}
\end{equation}
where $_2F_1$ is the hypergeometric function and $\Gamma$ the gamma function. This result is equivalent to the result obtained originally by \cite{1995Lokas} (see their Eq.~(38)). 

For the more general case of a generic linear matter power spectrum, we resort to numerical integration of Eq.~\eqref{skweness_final}. For the results shown in this paper (e.g. in Fig.~\ref{fig:cumulants}), we compute the linear power spectrum with {\sc camb} and for the cosmological parameters used in the Quijote simulation.

\subsection{$U_3 = \langle J_1J_2\rangle/\sigma_0$}

We now compute $U_3=\langle J_1J_2 \rangle/\sigma_0$.
With the previous notations, we notice that $\langle J_1J_2\rangle = \langle \nu(\nu^2-3I_2)\rangle =\langle \nu^3 \rangle - 3\langle \nu I_2 \rangle$, where the first term is the previously computed skewness. This leaves only $\langle \nu I_2 \rangle = \langle I_1 I_2 \rangle$ to be computed. Thanks to isotropy, this latter can be written as

\begin{equation}
\begin{split}
    \langle I_1I_2 \rangle &= 3\langle \delta \phi_{11}\phi_{22} \rangle - 3\langle \delta \phi_{12}^2 \rangle \\
    &= 3\langle (\phi_{11}+\phi_{22}+\phi_{33}) \phi_{11}\phi_{22} \rangle - 3\langle (\phi_{11}+\phi_{22}+\phi_{33}) \phi_{12}^2 \rangle \\
    &= 3(2\langle  \phi_{11}^2\phi_{22} \rangle +\langle  \phi_{11}\phi_{22}\phi_{33} \rangle-2\langle \phi_{11}\phi_{12}^2 \rangle - \langle \phi_{33 }\phi_{12}^2 \rangle).
\end{split}
\end{equation}

At leading order in standard perturbation theory (SPT), this reads
\begin{equation}
    \langle I_1 I_2 \rangle = 3\left(2\langle \phi_{11}^{(1)2} \phi_{22}^{(2)} \rangle  + 4 \langle \phi_{11}^{(2)} \phi_{11}^{(1)} \phi_{22}^{(1)} \rangle \right) + 3\langle \phi_{11}^{(2)}\phi_{22}^{(1)}\phi_{33}^{(1)} \rangle  - 2\left( 2\langle \phi_{11}^{(1)}\phi_{12}^{(1)}\phi_{12}^{(2)} \rangle + \langle \phi_{11}^{(2)}\phi_{12}^{(1)2} \rangle \right) - \left( 2\langle \phi_{33}^{(1)}\phi_{12}^{(1)}\phi_{12}^{(2)} \rangle + \langle \phi_{33}^{(2)}\phi_{12}^{(1)2} \rangle \right).
    \label{formal_I1I2}
\end{equation}

The computation of each of the above ensemble averages can be performed in a similar manner to the case of the skewness. Indeed, using a Fourier representation, the Poisson equation, and Wick's theorem, one has to integrate terms of the generic form
\begin{equation}
    2\int {\rm d}^3\bm{k_1} \int {\rm d}^3\bm{k_2} P(k_1)P(k_2)F_2(\bm{k_1},\bm{k_2}) \frac{K}{(\bm{k_1+k_2})^2}W(k_1R_1)W(k_2R_1)W(|{\bm k_1}+{\bm k_2}|R_2),
\end{equation}
where (i) the easiest way to proceed is to distinguish between two scales $R_1$ and $R_2$ as an intermediate step while in the end they will both be taken equal to our smoothing scale $R$ and (ii) $K$ is a polynomial in $\bm{k_1}$ and $\bm{k_2}$, which depends on which second derivatives of the gravitational potential are considered.

The term $1/(\bm{k_1} + \bm{k_2})^2$ unfortunately prevents direct analytical integration of the angular part of the previous form but differentiation by $R_2^2$ under the integral sign (sometimes referred to as \textit{Feynman's trick}) turns out to be a viable solution. Applied to the first term of Eq.~\eqref{formal_I1I2}, for example, we obtain
\begin{equation}
\begin{split}
    &\frac{\partial \langle \phi_{11}^2\phi_{22} \rangle}{\partial R_2^2} =-\frac{1}{2}(\bm{k_1+k_2})^2 \langle \phi_{11}^2\phi_{22} \rangle \\
   & = -\int {\rm d}^3\bm{k_1} \int {\rm d}^3\bm{k_2} P(k_1)P(k_2)F_2(\bm{k_1},\bm{k_2}) \frac{k_{1(1)}^2\left(k_{2(1)}^2 (k_{1}+k_{2})_{(2)}^2 + 2k_{2(2)}^2(k_{1}+k_{2})_{(1)}^2\right)}{k_1^2 k_2^2 }  W(k_1R_1)W(k_2R_1)W(|{\bm k_1}+{\bm k_2}|R_2), \label{cum2_terme1}
\end{split}
\end{equation}
where $k_{i(j)}$ is the $j$-th Cartesian component of ${\bm k_i}$ for $i \in \{1,2\}$. This form is much more suited to integration. Using the same decomposition of the Gaussian filter and the perturbation theory kernel into the basis of Legendre polynomials as in the skewness computation, we perform integration on the angle between ${\bm k_1}$ and ${\bm k_2}$, and a final integration on $R_2^2$. We finally obtain
\begin{equation}
\begin{split}
\label{eq:I1I2}
    \langle I_1 I_2 \rangle &= \int \frac{1}{70 R^4 (k_1 k_2 R^2)^{5/2}} \pi ^2 P(k_1) P(k_2) \sqrt{\frac{1}{k_1 k_2 R^2}} e^{-\frac{1}{2} R^2 (k_1^2+k_2^2)} \Biggl(-5 R^{10} (k_1^2-k_2^2)^4 (6 k_1^2-k_2^2) \biggl(\text{Ei}\bigl(-\frac{1}{2} (k_1-k_2)^2 R^2\bigr)-\text{Ei}\bigl(-\frac{1}{2} (k_1+k_2)^2 R^2\bigr)\biggr)\\
    & + e^{-\frac{1}{2} R^2 \left(k_1^2+k_2^2\right)} \biggl(-8 k_1 k_2 R^2 \left(R^2 \bigl(5 R^4 (k_1^2-k_2^2)^2 (6 k_1^2-k_2^2)-708 k_1^2-20 R^2 (6 k_1^4+19 k_1^2 k_2^2-k_2^4)-372 k_2^2\bigr)-960\right) \cosh (k_1 k_2 R^2)\\
    &-7680 \sinh (k_1 k_2 R^2)\biggr)-4 R^2 e^{-\frac{1}{2} R^2 \left(k_1^2+k_2^2\right)} \biggl(5 R^6 (k_1-k_2)^2 (k_1+k_2)^2 (6 k_1^2-k_2^2) (k_1^2+k_2^2)+1416 k_1^2+40 R^2 \left(6 k_1^4+35 k_1^2 k_2^2-k_2^4\right)\\
    &+2 R^4 \left(-30 k_1^6+301 k_1^4 k_2^2+84 k_1^2 k_2^4+5 k_2^6\right)+744 k_2^2\biggr) \sinh (k_1 k_2 R^2)\Biggr){\rm d}k_1 {\rm d}k_2,
\end{split}
\end{equation}
where $\text{Ei}$ is the exponential integral function $\text{Ei(z)} = -\int_z^\infty e^{-t}/tdt$.

Finally, we perform a numerical integration of Eq.~\eqref{eq:I1I2} with the Quijote linear power spectrum as for the skewness. Our results are displayed in Fig.~\ref{fig:cumulants}.

\subsection{$V_3 = \langle J_3 \rangle/\sigma_0$}

We now compute $V_3=\langle J_3\rangle/\sigma_0$. With the previous notations we note that $\langle J_3\rangle = \langle \nu^3 \rangle - {9}\langle I_1 I_2 \rangle/2 + {27}\langle I_3 \rangle$/2, where the first two terms are computed as part of $S_3$ and $U_3$ in the above subsections. The third term can then be written as
\begin{equation}
    \langle I_3 \rangle = \langle \phi_{11}\phi_{22}\phi_{33}\rangle+2\langle\phi_{12}\phi_{23}\phi_{13}\rangle-3\langle\phi_{11}\phi_{23}^2 \rangle,
\end{equation}
which, using isotropy and at leading order in SPT, can be rewritten as
\begin{equation}
    \langle I_3 \rangle = 3\langle \phi_{11}^{(2)}\phi_{22}^{(1)}\phi_{33}^{(1)}\rangle + 6\langle \phi_{12}^{(2)}\phi_{23}^{(1)}\phi_{13}^{(1)} \rangle - 3\left(2\langle \phi_{11}^{(1)}\phi_{23}^{(1)}\phi_{23}^{(2)} \rangle + \langle \phi_{11}^{(2)}(\phi_{23}^{(1)})^2 \rangle \right).
    \label{formal_I3}
\end{equation}

The exact same steps as in the computation of $\langle I_1I_2 \rangle$ are then performed to evaluate every ensemble average appearing in the previous Eq.~\eqref{formal_I3}, most notably the differentiation under the integral sign and the decomposition in Legendre polynomials of the Gaussian and perturbation theory kernels. We obtain 
\begin{equation}
\begin{split}
\label{eq:I3}
    \langle I_3 \rangle &= \int \frac{1}{140 (k_1 k_2 R^2)^{7/2}}\pi ^2 k_1 k_2 P(k_1) P(k_2) (k_1-k_2) (k_1+k_2) \sqrt{\frac{1}{k_1 k_2 R^2}} e^{-\frac{1}{2} R^2 (k_1^2+k_2^2)} \Biggl(-5 R^8 (k_1^2-k_2^2)^4 \text{Ei}\left(-\frac{1}{2} (k_1-k_2)^2 R^2\right)\\
    &+5 R^8 (k_1^2-k_2^2)^4 \text{Ei}\left(-\frac{1}{2} (k_1+k_2)^2 R^2\right)+e^{-\frac{1}{2} R^2 \left(k_1^2+k_2^2\right)} \biggl(8 k_1 k_2 R^2 \left(-5 R^4 (k_1^2-k_2^2)^2+20 R^2 (k_1^2+k_2^2)+48\right) \cosh (k_1 k_2 R^2)\\
    &-4 \left(5 R^6 (k_1^2-k_2^2)^2 (k_1^2+k_2^2)+40 R^2 (k_1^2+k_2^2)-2 R^4 \bigl(5 k_1^4-26 k_1^2 k_2^2+5 k_2^4\bigr)+96\right) \sinh (k_1 k_2 R^2)\biggr)\Biggr){\rm d}k_1 {\rm d}k_2,
\end{split}
\end{equation}
where $\text{Ei}$ is the exponential integral function $\text{Ei(z)} = -\int_z^\infty e^{-t}/tdt$. 

Again, we perform a numerical integration of Eq.~\eqref{eq:I3} with the Quijote linear power spectrum as for the skewness and $\langle I_1I_2 \rangle$. Our results are displayed in Fig.~\ref{fig:cumulants}.

\section{Behaviour of the Gram-Charlier}
\begin{figure}[h]
    \centering
    \includegraphics[width=\textwidth]{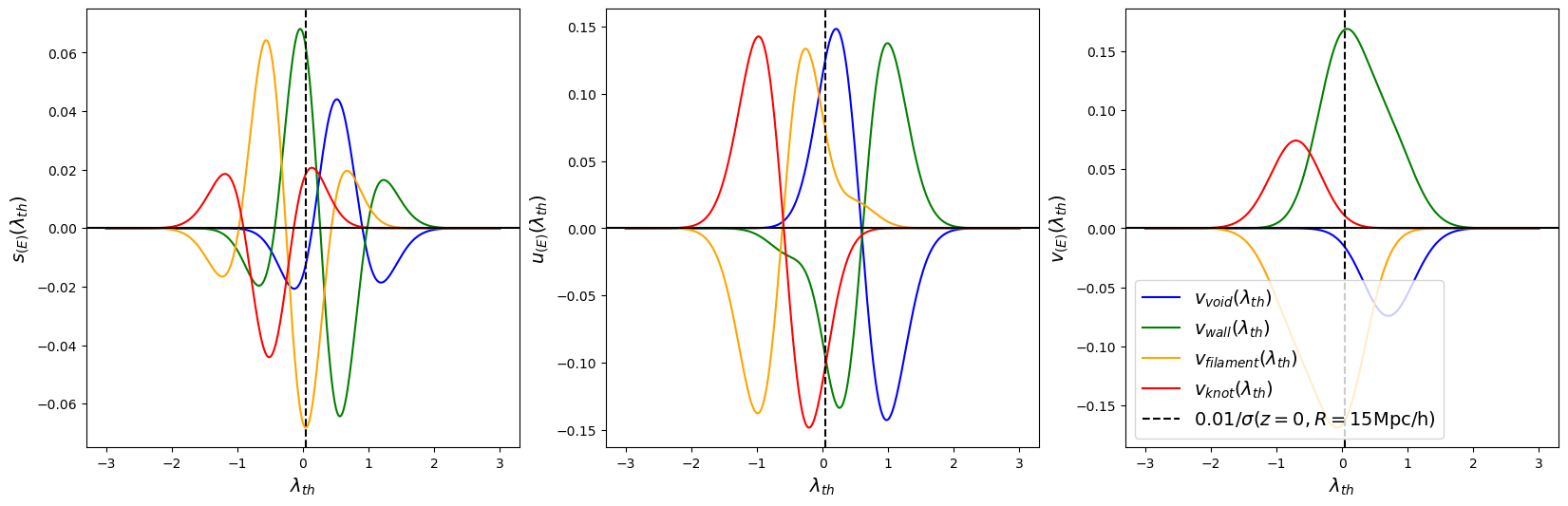}
    \caption{$s_{(E)}(\lambda_\text{th})$, $u_{(E)}(\lambda_\text{th})$, and $v_{(E)}(\lambda_\text{th})$ as a function of threshold predicted by the Gram-Charlier development. Results are shown for the four different environments: voids in blue, walls in green, filaments in orange, and knots in red. The vertical black line is the threshold used in the paper.}
    \label{fig:suv}
\end{figure}

The NLO prediction for the cosmic web abundances are derived in Eq.~\eqref{eq:gc_env}, which involves three functions of the threshold per environment:
\begin{equation}
    P_{(E)}(J_1, J_2, J_3) = P_{G,(E)}(J_1, J_2, J_3)+\sigma_0\bigl[s_{(E)}(\lambda_{\rm th})S_3+u_{(E)}(\lambda_{th,})U_3+v_{(E)}(\lambda_{\rm th})V_3\bigr].
\end{equation}
For the sake of completeness, in Fig.~\ref{fig:suv} we show the behaviour of these functions for the voids, walls, filaments, and knots in blue, green, yellow, and red, respectively, and as a function of the chosen threshold. The threshold chosen in our main analysis is the dashed black vertical line. We note that for a threshold of between $-2$ and $2$, the $s, \, u,$ and $v$ functions fluctuate a lot, thus potentially strongly modulating the environment probabilities depending on the values of the cumulants ---that is depending on how non-Gaussian the field is--- and most importantly on the value of the chosen threshold. This emphasises the importance of the choice of the threshold value.

\end{appendix}
\end{document}